\documentstyle[preprint,eqsecnum,aps]{revtex}

\def\lesssim{\ \raise.3ex\hbox{$<$}\kern-0.8em\lower.7ex\hbox{$\sim$}\ }
\def\gesim{\ \raise.3ex\hbox{$>$}\kern-0.8em\lower.7ex\hbox{$\sim$}\ }

\font\scripti=cmmi7
\font\scriptscripti=cmmi5
\def\sib#1{\setbox0 = \hbox{\scripti #1}
  \kern-.02em\copy0\kern-\wd0
  \kern.04em\box0} 
\def\ssib#1{\setbox0 = \hbox{\scriptscripti #1}
  \kern-.02em\copy0\kern-\wd0
  \kern.04em\box0} 

\font\tenib=cmmib10 
\skewchar\tenib='177 \skewchar\tenib='177 \skewchar\tenib='177
\def\ib{\fam10}

\def\pbold#1{\setbox0 = \hbox{$ #1 $}
  \kern-.022em\copy0\kern-\wd0
  \kern.011em\copy0\kern-\wd0
  \kern.011em\copy0\kern-\wd0
  \kern.011em\copy0\kern-\wd0
  \kern.011em\box0} 

\newcommand{\mib}{\ib}

\begin{document}
\draft
\title{Bound State and Order Parameter Mixing Effect by Nonmagnetic Impurity Scattering in Two-band Superconductors}
\author{Y. Ohashi}
\address{Institute of Physics, University of Tsukuba, Ibaraki 305, Japan}
\maketitle
\begin{abstract}
We investigate nonmagnetic impurity effects in two-band
superconductors, focusing on the effects of interband
scatterings. Within the Born approximation, it is known that interband
scatterings mix order parameters in the two bands. In particular, only
one averaged energy gap appears in the excitation spectrum in the
dirty limit. [G. Gusman: J. Phys. Chem. Solids {\bf 28} (1967) 2327.]
In this paper, we take into account the interband scattering within
the $t$-matrix approximation beyond the Born approximation in the
previous work. We show that, although the interband scattering is
responsible for the mixing effect, this effect becomes weak when the
interband scattering becomes very strong. In the strong interband
scattering limit, a two-gap structure corresponding to two order
parameters recovers in the superconducting density of states. We also
show that a bound state appears around a nonmagnetic impurity
depending on the phase of interband scattering potential.
\end{abstract}
\vskip5mm
\narrowtext

\sloppy
\maketitle
\newpage
\section{Introduction}
Since the discovery of superconductivity in MgB$_2$,\cite{Akimitsu} great efforts have been done on this material in order to clarify the character of this 40[K]-superconductor. Experiments on the specific heat\cite{Bou} and penetration depth\cite{Man} indicate a finite energy gap below the superconducting transition temperature $T_{\rm c}$. The presence of a Boron-isotope effect\cite{Bud} implies that phonon plays a crucial role in the pairing interaction. However, the origin of this superconductivity is still in controversial and various mechanism have been proposed.\cite{Yamaji,Furukawa,Imada}
\par
Besides the mechanism, possibility of multi-gap superconductivity has been discussed in this material.\cite{Shulga,Liu,Golubov,Nakai} Indeed, a two-gap structure has been directly observed by tunnelling experiments.\cite{Giu,Nai} The temperature dependence of the specific heat also indicates the presence of two energy gaps.\cite{Bou}
Since MgB$_2$ has two distinct Fermi surfaces originating from $\sigma$- and $\pi$-orbitals of boron atoms,\cite{Kortus} it is probable that the superconducting order parameters are different in the two bands. We also mention that multi-gap superconductivity has been investigated also in the superconducting state of transition metals.\cite{Shul,Kondo,Soda,Legget,Gusman,Chow1,Chow2,Tang,Sung} 
\par
When the order parameters are different among bands, it is known that impurities with interband scatterings strongly affect the superconducting state even if impurities are nonmagnetic.\cite{Gusman,Chow1,Chow2,Tang,Sung} In particular, Gusman\cite{Gusman} showed within the Born approximation that the order parameters in a two-band superconductor are mixed with each other by interband scatterings. In the dirty limit, only one averaged energy gap appears in the superconducting density of states. Although this impurity effect was originally considered for the superconductivity in transition metals, this effect can be expected also in MgB$_2$ if this superconductor really has two distinct order parameters in the two bands. 
\par
So far, the mixing of order parameters by interband scatterings has been discussed within the Born approximation.\cite{Gusman,Chow1,Chow2,Tang,Sung} With this regard, we mention that higher order scattering processes beyond the Born approximation sometimes lead to qualitatively new effects in the case when the impurity scattering affects superconductivity. (For example, formation of a bound state around a magnetic impurity in $s$-wave superconductivity.\cite{Shiba}) Thus it is expected also in the present case that higher order scattering processes may cause new phenomena which cannot be described within the Born approximation. 
\par
In this paper we investigate nonmagnetic impurity effects in a model two-band superconductor with two different order parameters. We focus on the mixing of the order parameters by interband scatterings. In contrast to the previous works which employ the Born approximation, we take into account multi-scattering processes in terms of the inter- and intra-band scatterings using the $t$-matrix approximation. We clarify how the mixing effect of the order parameters which has been discussed within the Born approximation is modified by higher order scattering processes. We show that, although the interband scatterings are crucial for the mixing effect, at the same time, this effect is suppressed when the interband scattering is very strong; this result is not obtained within the Born approximation. As another new result by the $t$-matrix approximation, we also show that a bound state is formed  around a nonmagnetic impurity depending on the detailed character of impurity scattering. 
\par
This paper is organized as follows: After explaining our model in $\S$2, we consider a single impurity problem and discuss a bound state in $\S$3. In $\S$4, we investigate the case of finite impurity concentration, which is followed by summary in $\S$5.
\par
\vskip3mm
\section{Model Two-band Superconductor}
\subsection{Hamiltonian}
We consider a three-dimensional two-band $s$-wave superconductor. We refer to the two bands as $a$- and $b$-band. The Hamiltonian is given by\cite{Shul,Kondo,Soda,Legget,Gusman}
\begin{eqnarray}
H=
\sum_{{\sib p},\sigma}\varepsilon^a_{\sib p}
a^\dagger_{{\sib p}\sigma}a_{{\sib p}\sigma}
+
\sum_{{\sib p},\sigma}\varepsilon^b_{\sib p}
b^\dagger_{{\sib p}\sigma}b_{{\sib p}\sigma}
&-&
g_{aa}\sum_{{\sib p},{\sib p}'}
a_{{\sib p}\uparrow}^\dagger a_{-{\sib p}\downarrow}^\dagger
a_{-{\sib p}'\downarrow} a_{{\sib p}'\uparrow}
-
g_{bb}\sum_{{\sib p},{\sib p}'}
b_{{\sib p}\uparrow}^\dagger b_{-{\sib p}\downarrow}^\dagger
b_{-{\sib p}'\downarrow} b_{{\sib p}'\uparrow}
\nonumber
\\
&-&
g_{ab}\sum_{{\sib p},{\sib p}'}
a_{{\sib p}\uparrow}^\dagger a_{-{\sib p}\downarrow}^\dagger
b_{-{\sib p}'\downarrow} b_{{\sib p}'\uparrow}
-
g_{ba}\sum_{{\sib p},{\sib p}'}
b_{{\sib p}\uparrow}^\dagger b_{-{\sib p}\downarrow}^\dagger
a_{-{\sib p}'\downarrow} a_{{\sib p}'\uparrow}
\nonumber
\\
&+&
H_{\rm imp}.
\label{eq.2.1}
\end{eqnarray}
Here, $a_{{\sib p}\sigma}$ ($b_{{\sib p}\sigma}$) represents the annihilation operator of an electron in the $a$-band ($b$-band). The kinetic energies of the $a$- and $b$-band measured from the Fermi level are, respectively, given by $\varepsilon^a_{\sib p}$  and $\varepsilon^b_{\sib p}$. In this paper, we do not take into account the detailed band structure of MgB$_2$ and simply assume two isotropic three-dimensional bands. The $a$-band ($b$-band) has an intraband pairing interaction described by $g_{aa}$ ($g_{bb}$). In addition, eq. (\ref{eq.2.1}) also includes an interband pairing interaction described by $g_{ab}$ and $g_{ba}~(=g_{ab}^*)$. We do not specify the origin of the pairing interactions and, for simplicity, assume the same cut-off energy $\omega_{\rm D}$ for all the pairing interactions. We briefly mention that different cut-off energies between the two bands are crucial in considering the isotope effect as discussed by Kondo.\cite{Kondo} The last term in eq. (\ref{eq.2.1}) represents the nonmagnetic impurity scattering: 
\begin{eqnarray}
H_{\rm imp}=
\sum_{{\sib R}_i,{\sib p},{\sib p}',\sigma}
{\rm e}^{{\rm i}({\sib p}-{\sib p}')\cdot{\sib R}_i}
\bigl[
v_{aa}a_{{\sib p}\sigma}^\dagger a_{{\sib p}'\sigma}
+v_{bb}b_{{\sib p}\sigma}^\dagger b_{{\sib p}'\sigma}
+v_{ab}a_{{\sib p}\sigma}^\dagger b_{{\sib p}'\sigma}
+v_{ba}b_{{\sib p}\sigma}^\dagger a_{{\sib p}'\sigma}
\bigr],
\label{eq.2.2}
\end{eqnarray}
where $v_{aa}$ and $v_{bb}$, respectively, describe intraband impurity scattering in the $a$- and $b$-band while $v_{ab}$ and $v_{ba}~(=v_{ab}^*)$ represent interband scattering between the two bands. In eq. (\ref{eq.2.2}), we have neglected momentum dependence of scattering potentials for simplicity and retained the $s$-wave scattering component only. In the case of a single impurity problem, we put the position of the impurity ${\mib R}_i=0$ and drop the summation in terms of ${\mib R}_i$ in eq. (\ref{eq.2.2}).
\par
We employ the mean field approximation and introduce the two superconducting order parameters in the $a$- and $b$-band:
\begin{eqnarray}
\left\{
\begin{array}{l}
\Delta_a=-g_{aa}\sum_{\sib p}
\langle 
a_{{\sib p}\uparrow}a_{-{\sib p}\downarrow} 
\rangle, \\
\Delta_b=-g_{bb}\sum_{\sib p}
\langle 
b_{{\sib p}\uparrow}b_{-{\sib p}\downarrow} 
\rangle. \\
\end{array}
\right.
\label{eq.2.3}
\end{eqnarray}
Then the mean field Hamiltonian of eq. (\ref{eq.2.1}) is obtained as (we drop constant terms)
\begin{equation}
H_{\rm MF}=
\sum_{\sib p}\Psi_{\sib p}^\dagger(\varepsilon^a_{\sib p}\rho_3-{\hat \Delta}_a)\Psi_{\sib p}
+
\sum_{\sib p}\Phi_{\sib p}^\dagger(\varepsilon^b_{\sib p}\rho_3-{\hat \Delta}_b)\Phi_{\sib p}
+
H_{\rm imp},
\label{eq.2.4}
\end{equation}
where 
$\Psi_{\sib p}^\dagger\equiv(a^\dagger_{{\sib p}\uparrow},a_{-{\sib p}\downarrow})$, and $\Phi_{\sib p}^\dagger\equiv(b^\dagger_{{\sib p}\uparrow},b_{-{\sib p}\downarrow})$ represent two-component Nambu field operators in the $a$- and $b$-band, respectively; $\rho_i$ $(i=1,2,3)$ are corresponding Pauli matrices. In eq. (\ref{eq.2.4}), ${\hat \Delta}_j$ ($j=a,b$) are given by
\begin{eqnarray}
{\hat \Delta}_j=
\left(
\begin{array}{cc}
0 & {\tilde \Delta}_j \\
{\tilde \Delta}_j^* & 0 \\
\end{array}
\right)
=
{\rm Re}[{\tilde \Delta}_j]\rho_1-{\rm Im}[{\tilde \Delta}_j]\rho_2,\label{eq.gap1}
\end{eqnarray} 
where ${\tilde \Delta}_j$ are related to the order parameters as
\begin{eqnarray}
\left\{
\begin{array}{l}
\displaystyle
{\tilde \Delta}_a=\Delta_a+{g_{ab} \over g_{bb}}\Delta_b, \\
\displaystyle
{\tilde \Delta}_b=\Delta_b+{g_{ba} \over g_{aa}}\Delta_a. \\
\end{array}
\right.
\label{eq.2.5}
\end{eqnarray}
The impurity term $H_{\rm imp}$ under the Nambu representation is given by
\begin{equation}
H_{\rm imp}=
\sum_{{\sib R}_i,{\sib p},{\sib p}'}
{\rm e}^{{\rm i}({\sib p}-{\sib p}')\cdot{\sib R}_i}
\Bigl[
\Psi^\dagger_{\sib p}{\hat v}_{aa}\Psi_{{\sib p}'}
+
\Phi^\dagger_{\sib p}{\hat v}_{bb}\Phi_{{\sib p}'}
+
\Psi^\dagger_{\sib p}{\hat v}_{ab}\Phi_{{\sib p}'}
+
\Phi^\dagger_{\sib p}{\hat v}_{ba}\Psi_{{\sib p}'}
\Bigr].
\label{eq.2.6}
\end{equation}
Here the matrix impurity potentials are defined by ${\hat v}_{aa}=v_{aa}\rho_3$, ${\hat v}_{bb}=v_{bb}\rho_3$ and 
\begin{eqnarray}
{\hat v}_{ab}=
\left(
\begin{array}{cc}
v_{ab} & 0 \\
0 & -v_{ab}^*
\end{array}
\right)
={\rm Re}[v_{ab}]\rho_3+i{\rm Im}[v_{ab}]{\bf 1},~~~~~~{\hat v}_{ba}={\hat v}_{ab}^\dagger.
\label{eq.vab}
\end{eqnarray}
\par
We can take $\Delta_a$ real and $g_{ab}=g_{ba}>0$ with use of a gauge transformation. Complex quantities in our model are then $\Delta_b=|\Delta_b|{\rm e}^{{\rm i}\phi_b}$ and $v_{ab}=|v_{ab}|{\rm e}^{{\rm i}\theta_{ab}}$. 
\vskip3mm
\subsection{Clean system}
\vskip2mm
In this subsection, we briefly review the theory of two-band superconductivity in the clean system\cite{Shul,Kondo,Soda,Legget} in order to see how the impurity scatterings modify the superconducting state in later sections. The one-particle matrix thermal Green function in the $j$-band ($j=a,b$) is given by
\begin{equation}
G^0_j({\mib p},{\rm i}\omega_m)={1 \over {\rm i}\omega_m-\varepsilon^j_{\sib p}\rho_3+{\hat \Delta}_j}.
\label{eq.2.7}
\end{equation}
Here $\omega_m$ is the Fermion Matsubara frequency. When we calculate the superconducting density of states (SDOS) in the $j$-band, we obtain 
\begin{equation}
N_j(\omega)=-{1 \over \pi}
\sum_{\sib p}
{\rm Im}G^0_j({\mib p},i\omega_m\to\omega+i\delta)
=N_j(0){\rm Re}
\bigl[
{\omega \over \sqrt{\omega^2-|{\tilde \Delta}_j|^2}}
\bigr],
\label{eq.2.8}
\end{equation}
where $N_j(0)$ is the density of states at the Fermi level in the normal state. Equation (\ref{eq.2.8}) shows that the energy gap in SDOS is governed by, not the order parameter $\Delta_j$ itself, but ${\tilde \Delta}_j$ defined in eq. (\ref{eq.2.5}). Since ${\tilde \Delta}_j$ includes both $\Delta_a$ and $\Delta_b$, the superconducting states in the two bands couple with each other by the interband pairing interaction $g_{ab}$.
\par
Substituting eq. (\ref{eq.2.7}) into the gap equation $\Delta_j=g_{jj}T\sum_{{\sib p},\omega_m}G^0_j({\mib p},{\rm i}\omega_m)_{12}$, we obtain
\begin{equation}
\Delta_j=g_{jj}N_j(0)
\int_0^{\omega_{\rm D}}{\rm d}\varepsilon
{
{\tilde \Delta}_j 
\over 
\sqrt{\varepsilon^2+|{\tilde \Delta}_j|^2}
}
\tanh{\beta \over 2}
\sqrt{\varepsilon^2+|{\tilde \Delta}_j|^2}~~~~~~(j=a,b),
\label{eq.2.9}
\end{equation}
where $\beta=1/T$ is the inverse of temperature. Substituting eq. (\ref{eq.2.5}) into eq. (\ref{eq.2.9}), we find $\Delta_b=C\Delta_a$, where $C$ is given by
\begin{equation}
C=
{
1-g_{aa}N_a(0)
\int_0^{\omega_{\rm D}}{\rm d}\varepsilon
{1 \over \sqrt{\varepsilon^2+|{\tilde \Delta}_a|^2}}
\tanh{\beta \over 2}
\sqrt{\varepsilon^2+|{\tilde \Delta}_a|^2}
\over 
g_{aa}{g_{ab} \over g_{bb}}N_a(0)
\int_0^{\omega_{\rm D}}{\rm d}\varepsilon
{1 \over \sqrt{\varepsilon^2+|{\tilde \Delta}_a|^2}}
\tanh{\beta \over 2}
\sqrt{\varepsilon^2+|{\tilde \Delta}_a|^2}
}.
\label{eq.2.10}
\end{equation}
Thus when $\Delta_a$ becomes finite, $\Delta_b$ also becomes finite as far as $g_{ab}\ne 0$ and $g_{bb}\ne 0$.\cite{Shul} We also find that $\Delta_b$ can be also taken real as well as $\Delta_a$ in the clean system.        
\par
The superconducting transition temperature $T_{\rm c}$ is determined by\cite{Shul,Soda}
\begin{equation}
T_{\rm c}={2\gamma\omega_{\rm D} \over \pi}
{\rm exp}
\Bigl[
-
{
{\bar g}_{aa}+{\bar g}_{bb}-
\sqrt{({\bar g}_{aa}-{\bar g}_{bb})^2+4{\bar g}_{ab}{\bar g}_{ba}}
\over 
2({\bar g}_{aa}{\bar g}_{bb}-{\bar g}_{ab}{\bar g}_{ba})
}
\Bigr]~~~~(\gamma=1.78),
\label{eq.2.11}
\end{equation}
where ${\bar g}_{aa}\equiv g_{aa}N_a(0)$, ${\bar g}_{bb}\equiv g_{bb}N_a(0)$, ${\bar g}_{ab}\equiv g_{ab}N_b(0)$ and ${\bar g}_{ba}\equiv g_{ba}N_a(0)$. 
\par
Temperature dependence of the order parameters in the clean system is shown in Fig. 1. When ${\bar g}_{ab}\sim({\bar g}_{aa},~{\bar g}_{bb}$) (Fig. 1(a)), the overall temperature dependence is similar to the temperature dependence of the order parameter in an ordinary single-band BCS superconductor. In weak coupling BCS theory, the magnitude of the energy gap at $T=0$ is related to $T_{\rm c}$ as $2\Delta_{\rm BCS}(T=0)/T_{\rm c}=3.54$. On the other hand, as proved by Soda and Wada,\cite{Soda} $2{\tilde \Delta}_a(T=0)/T_{\rm c}$ is larger than 3.54 while $2{\tilde \Delta}_b(T=0)/T_{\rm c}$ is smaller than 3.54 in Fig. 1(a). When ${\bar g}_{ab}\ll({\bar g}_{aa},~{\bar g}_{bb})$ (Fig. 1(b)), the smaller order parameter ($\Delta_b$) shows a curious temperature dependence:\cite{Shul} When ${\bar g}_{ab}=0$, the $b$-band remains in the normal state down to $T/T_{\rm c}=0.513$ (where $T_{\rm c}$ is the transition temperature in the case of ${\bar g}_{ab}=0.002$). Thus $\Delta_b$ between $0.513<T/T_{\rm c}<1$ in Fig. 1(b) is due to a proximity effect caused by the interband pairing interaction $g_{ab}$. This induced order parameter appears even if the intraband interaction in the $b$-band is {\it repulsive} ($g_{bb}<0$). We shows an example in Fig. 1(c).
\vskip3mm
\section{Bound State around a Single Impurity in Two-band Superconductor}
\vskip2mm
\subsection{Equation of bound state}
\vskip3mm
Let us consider the electronic state around an impurity at ${\mib R}_i=0$. In this case, within the neglect of the spatial variation of the order parameter, the one-particle matrix thermal Green function in the $a$-band $G_a({\mib p},{\mib p}',i\omega_m)$ has the form
\begin{equation}
G_a({\mib p},{\mib p}',i\omega_m)=
G_a^0({\mib p},i\omega_m)\delta_{{\sib p},{\sib p}'}
+
G_a^0({\mib p},i\omega_m){\hat t}_{aa}({\rm i}\omega)
G_a^0({\mib p}',i\omega_m).
\label{eq.3.1}
\end{equation}
Here the $t$-matrix ${\hat t}_{aa}$ obeys
\begin{eqnarray}
\left\{
\begin{array}{l}
{\hat t}_{aa}({\rm i}\omega_m)=
{\hat v}_{aa}+
{\hat v}_{aa}\sum_{\sib p}
G_a^0({\mib p},{\rm i}\omega_m){\hat t}_{aa}({\rm i}\omega_m)
+
{\hat v}_{ab}\sum_{\sib p}
G_b^0({\mib p},{\rm i}\omega_m){\hat t}_{ba}({\rm i}\omega_m),\\
{\hat t}_{ba}({\rm i}\omega_m)=
{\hat v}_{ba}+
{\hat v}_{ba}\sum_{\sib p}
G_a^0({\mib p},{\rm i}\omega_m){\hat t}_{aa}({\rm i}\omega_m)
+
{\hat v}_{bb}\sum_{\sib p}
G_b^0({\mib p},{\rm i}\omega_m){\hat t}_{ba}({\rm i}\omega_m).
\end{array}
\right.
\label{eq.3.2}
\end{eqnarray}
The solution of eq.(\ref{eq.3.2}) is given by
\begin{equation}
{\hat t}_{aa}({\rm i}\omega_m)
=
[1-{\hat V}_{aa}({\rm i}\omega_m)\sum_{\sib p}G^0_{aa}({\mib p},{\rm i}\omega_m)]^{-1}
{\hat V}_{aa}({\rm i}\omega_m),
\label{eq.3.3}
\end{equation}
where the effective impurity potential in the $a$-band defined by ${\hat V}_{aa}({\rm i}\omega_m)$ includes interband scatterings between the two bands as 
\begin{equation}
{\hat V}_{aa}({\rm i}\omega_m)=
{\hat v}_{aa}
+
{\hat v}_{ab}
\sum_{\sib p}
G_b^0({\mib p},{\rm i}\omega_m)
[1-{\hat v}_{bb}\sum_{\sib p}G_b^0({\mib p},{\rm i}\omega_m)]^{-1}
{\hat v}_{ba}.
\label{eq.3.4}
\end{equation}
\par
A bound state caused by the impurity scattering is obtained as a pole of the analytic-continued $t$-matrix of eq. (\ref{eq.3.3}). The equation of the bound state energy $\omega$ is then given by ($\omega<{\rm Min}[|{\tilde \Delta}_a|,|{\tilde \Delta}_b|]$)
\begin{eqnarray}
{
\omega^2-\cos(2\theta_{ab}){\tilde \Delta}_a{\tilde \Delta}_b
\over
\sqrt{{\tilde \Delta}_a^2-\omega^2}
\sqrt{{\tilde \Delta}_b^2-\omega^2}}
={1 \over 2|{\bar v}_{ab}|^2}(1+\alpha_a^2+\alpha_b^2+(|{\bar v}_{ab}|^2-\alpha_a\alpha_b)^2),
\label{eq.3.5}
\end{eqnarray}
where $\alpha_a\equiv\pi N_a(0)v_{aa}$ and $\alpha_b\equiv\pi N_b(0)v_{bb}$ represent intraband impurity potentials normalized by the density of states; ${\bar v}_{ab}\equiv\pi\sqrt{N_a(0)N_b(0)}v_{ab}=|{\bar v}_{ab}|{\rm e}^{{\rm i}\theta_{ab}}$ . Since we consider a single impurity problem and neglect spatial variation of the order parameters, we have taken both ${\tilde \Delta}_a$ and ${\tilde \Delta}_b$ real in eq. (\ref{eq.3.5}). The derivation of eq. (\ref{eq.3.5}) is summarized in Appendix A.
\par
\vskip2mm
\subsection{Conditions for the existence of bound state}
\vskip3mm
Since the present theory is symmetric in terms of the $a$- and $b$-band, we may consider only the case of $|{\tilde \Delta}_a|\ge|{\tilde \Delta}_b|$ without loss of generality. The left hand side (LHS) in eq. (\ref{eq.3.5}) equals $-\cos(2\theta_{ab}){\rm sign}[{\tilde \Delta}_a{\tilde \Delta}_b]$ at $\omega=0$ while it becomes $\pm\infty$ at $\omega=|{\tilde \Delta}_b|$ depending on the sign of ${\tilde \Delta}_b^2-\cos(2\theta_{ab}){\tilde \Delta}_a{\tilde \Delta}_b$. We then find the following three cases about the formation of the bound state:
\begin{enumerate}
\item[(1)] $\cos(2\theta_{ab}){\rm sign}[{\tilde \Delta}_a{\tilde \Delta}_b]\le 0$: A bound state is obtained if the right hand side (RHS) in eq. (\ref{eq.3.5}) is larger than $|\cos(2\theta_{ab})|$. 
\par
\item[(2)] $\cos(2\theta_{ab}){\rm sign}[{\tilde \Delta}_a{\tilde \Delta}_b]>0$ and ${\tilde \Delta}_b^2-\cos(2\theta_{ab}){\tilde \Delta}_a{\tilde \Delta}_b>0$: A bound state exists because LHS in eq. (\ref{eq.3.5}) monotonously increases from $-\cos(2\theta_{ab})$ to $+\infty$ as $\omega$ changes from $0$ to $|{\tilde \Delta}_b|$. 
\par
\item[(3)] $\cos(2\theta_{ab}){\rm sign}[{\tilde \Delta}_a{\tilde \Delta}_b]>0$ and ${\tilde \Delta}_b^2-\cos(2\theta_{ab}){\tilde \Delta}_a{\tilde \Delta}_b\le 0$: A bound state is absent below $|{\tilde \Delta}_b|$ because LHS in eq. (\ref{eq.3.5}) is always negative. 
\end{enumerate} 
A bound state is not obtained when the interband potential scattering ${\bar v}_{ab}$ is absent because RHS in eq. (\ref{eq.3.5}) goes infinity. In addition, when ${\tilde \Delta}_a$ and ${\tilde \Delta}_b$ have the same sign and the interband potential scattering is real, one does not obtain a bound state. Thus, the phase of the interband potential scattering and the difference of the signs of ${\tilde \Delta}_a$ and ${\tilde \Delta}_b$ are crucial in obtaining the bound state. 
\par
Although the interband scattering is crucial for the bound state, eq. (\ref{eq.3.5}) also shows that the bound state disappears in the limit $|{\bar v}_{ab}|\to\infty$. (Note that RHS in eq. (\ref{eq.3.5}) goes infinity in this limit.) When we write the interband scattering term as $[v_{ab}a^\dagger_{0\sigma}b_{0\sigma}+h.c.]$ ($a^\dagger_{0\sigma}$ ($b_{0\sigma}$): creation (annihilation) operator of an electron in the $a$-band ($b$-band) at ${\mib R}=0$) and diagonalize this term, we obtain two energy levels $E_\pm=\pm|v_{ab}|$. In the limit $|v_{ab}|\to\infty$, the lower level $E_-$ is occupied by two electrons while the higher state $E_+$ becomes irrelevant owing to its extremely high energy. Then since electrons no longer come to the impurity site, the effect of the interband scattering disappears in the limit $|{\bar v}_{ab}|\to\infty$.
\par
Dependence of the bound state energy on $|{\bar v}_{ab}|$ is shown in Fig. 2. A bound state appears when the interband scattering becomes finite; however, the bound state energy again approaches the energy gap $|{\tilde \Delta}_b|$ when $|{\tilde v}_{ab}|\gg 1$ because of the reason explained above. In Fig. 1(a), the bound state energy increases as the phase of interband impurity scattering $\theta_{ab}$ decreases. In the case of Fig. 1(a) (${\tilde \Delta}_a=2$ and ${\tilde \Delta}_b=1$), the bound state disappears for $\theta_{ab}\ge\pi/6$ because the case-(3) discussed above is satisfied. The bound state energy also increases when the intraband scattering ($\alpha_b$) becomes strong, as shown in Fig. 1(b). This is because electrons cannot come to the impurity site when the intraband impurity potential is strong; thus the effect of the interband scattering is weakened by the intraband impurity scattering. We also find this effect in eq. (\ref{eq.3.5}); when $\alpha_a\to\infty$ and/or $\alpha_b\to\infty$, RHS in eq. (\ref{eq.3.5}) goes infinity as in the case of $|{\bar v}_{ab}|\to 0$.
\vskip3mm
\section{Order Parameter Mixing in Dirty Two-band Superconductors}
\vskip2mm
\subsection{Formulation: $t$-matrix approximation}
\vskip2mm
In this section, we consider the case of finite impurity concentration using the $t$-matrix approximation. After taking spatial average in terms of the distribution of impurities, we obtain the renormalized one-particle matrix thermal Green function in the $a$-band as
\begin{eqnarray}
G_a({\mib p},{\rm i}\omega_m)
&=&
{
1 
\over 
{\rm i}{\omega}_m-\varepsilon^a_{\sib p}\rho_3+
{\hat \Delta}_a-\Sigma_{aa}({\rm i}\omega_m)
}
\nonumber
\\
&\equiv&
{1 
\over 
{\rm i}{\bar \omega}_m^a-\varepsilon^a_{\sib p}\rho_3+
{\hat {\bar \Delta}}_a
}.
\label{eq.4.1}
\end{eqnarray}
The self-energy $\Sigma_{aa}({\rm i}\omega_m)$ within the $t$-matrix approximation is given by 
\begin{eqnarray}
\left\{
\begin{array}{l}
\Sigma_{aa}({\rm i}\omega_m)=n_{\rm imp}
[1-{\bar V}_{aa}({\rm i}\omega_m)\sum_{\sib p}G_a({\mib p},{\rm i}\omega_m)]^{-1}{\bar V}_{aa},
\\
{\bar V}_{aa}({\rm i}\omega_m)=
{\hat v}_{aa}+{\hat v}_{ab}\sum_{\sib p}
G_b({\mib p},{\rm i}\omega_m)
[1-{\hat v}_{bb}G_b({\mib p},{\rm i}\omega_m)]^{-1}
{\hat v}_{ba}.
\end{array}
\right.
\label{eq.4.2}
\end{eqnarray}
Here $n_{\rm imp}$ is the number density of impurities. In eq. (\ref{eq.4.2}), ${\bar V}_{aa}$ has the same structure as ${\hat V}_{aa}$ in eq. (\ref{eq.3.4}) except that the renormalized Green function is now used in place of $G_b^0$. The renormalized Green function in the $b$-band is also given by eqs. (\ref{eq.4.1}) and (\ref{eq.4.2}) where the indices '$a$' and '$b$' are interchanged. We mention that $\Sigma_{aa}$ has the same structure as ${\hat t}_{aa}$ in eq. (\ref{eq.3.3}).
\par
In Eq. (\ref{eq.4.1}), ${\bar \omega}_m^j$ is a renormalized Matsubara frequency and
\begin{eqnarray}
{\hat {\bar \Delta}}_j=
\left(
\begin{array}{cc}
0 & {\bar {\tilde \Delta}}_j \\
{\bar {\tilde \Delta}}_j^* & 0 \\
\end{array}
\right)
=
{\rm Re}[{\bar {\tilde \Delta}}_j]\rho_1-{\rm Im}[{\bar {\tilde \Delta}}_j]\rho_2
\label{eq.4.3}
\end{eqnarray}
with ${\bar {\tilde \Delta}}_j$ being a renormalized ${\tilde \Delta}_j$.\cite{foot1} Their expressions are derived in Appendix B and we only show the results here:
\begin{eqnarray}
\left\{
\begin{array}{l}
\displaystyle
{\rm i}{\bar \omega}^a_m={\rm i}\omega_m
+
{u_a \over D_a}
\bigl[
{|{\bar v}_{ab}|^2 \over 1+\alpha_b^2}
{{\rm i}{\bar \omega}_m^b \over {\bar \Omega}_b}
+
\eta_a
{{\rm i}{\bar \omega}_m^a \over {\bar \Omega}_a}
\bigr],
\\
\displaystyle
{\rm i}{\bar \omega}^b_m={\rm i}\omega_m
+
{u_b \over D_b}
\bigl[
{|{\bar v}_{ba}|^2 \over 1+\alpha_a^2}
{{\rm i}{\bar \omega}_m^a \over {\bar \Omega}_a}
+
\eta_b
{{\rm i}{\bar \omega}_m^b \over {\bar \Omega}_b}
\bigr],
\end{array}
\right.
\label{eq.4.4}
\end{eqnarray}
\begin{eqnarray}
\left\{
\begin{array}{l}
\displaystyle
{\bar {\tilde \Delta}}_a={\tilde \Delta}_a
+
{u_a \over D_a}
\bigl[
{{\bar v}_{ab}^2 \over 1+\alpha_b^2}
{{\bar {\tilde \Delta}}_b \over {\bar \Omega}_b}
+
\eta_a
{{\bar {\tilde \Delta}}_a \over {\bar \Omega}_a}
\bigr],
\\
\displaystyle
{\bar {\tilde \Delta}}_b={\tilde \Delta}_b
+
{u_b \over D_b}
\bigl[
{{\bar v}_{ba}^2 \over 1+\alpha_a^2}
{{\bar {\tilde \Delta}}_a \over {\bar \Omega}_a}
+
\eta_b
{{\bar {\tilde \Delta}}_b \over {\bar \Omega}_b}
\bigr],
\end{array}
\right.
\label{eq.4.5}
\end{eqnarray}
where ${\bar \Omega}_j=\sqrt{{\bar \omega}_m^j{}^2+|{\bar {\tilde \Delta}}_j|^2}$, $u_j=n_{\rm imp}/(\pi N_j(0))$ and 
\begin{eqnarray}
\left\{
\begin{array}{l}
\displaystyle
\eta_a=
{|{\bar v}_{ab}|^4 \over (1+\alpha_b^2)^2}
+
\bigl(
\alpha_a-{\alpha_b \over 1+\alpha_b^2}|{\bar v}_{ab}|^2
\bigr)^2,
\\
\displaystyle
\eta_b=
{|{\bar v}_{ba}|^4 \over (1+\alpha_a^2)^2}
+
\bigl(
\alpha_b-{\alpha_a \over 1+\alpha_a^2}|{\bar v}_{ba}|^2
\bigr)^2,
\end{array}
\right.
\label{eq.4.6}
\end{eqnarray}
\begin{eqnarray}
\left\{
\begin{array}{l}
\displaystyle
D_a=1+\eta_a+{2 \over 1+\alpha_b^2}
{1 \over {\bar \Omega}_a{\bar \Omega_b}}
\bigl[
{\bar \omega}_m^a{\bar \omega}_m^b|{\bar v}_{ab}|^2
+
{\rm Re}
[
{\bar v}_{ab}^2{\bar {\tilde \Delta}}^*_a{\bar {\tilde \Delta}}_b
]
\bigr],
\\
\displaystyle
D_b=1+\eta_b+{2 \over 1+\alpha_a^2}
{1 \over {\bar \Omega}_a{\bar \Omega_b}}
\bigl[
{\bar \omega}_m^a{\bar \omega}_m^b|{\bar v}_{ba}|^2
+
{\rm Re}
[
{\bar v}_{ba}^2{\bar {\tilde \Delta}}^*_b{\bar {\tilde \Delta}}_a
]
\bigr].
\end{array}
\right.
\label{eq.4.7}
\end{eqnarray}
\par
\vskip3mm
\subsection{Gap equation}
\vskip2mm
The order parameter in the $j$-band, $\Delta_j$, is determined by the gap equation $\Delta_j=g_{jj}T\sum_{{\sib p},\omega_m}G_j({\mib p},{\rm i}\omega_m)_{12}$:
\begin{equation}
\Delta_j={\bar g}_{jj}{\pi \over \beta}
\sum_{\omega_m}
{
{\bar {\tilde \Delta}}_j 
\over 
\sqrt{{\bar \omega}_m^j{}^2+|{\bar {\tilde \Delta}}_j|^2}
}
~~~~(j=a,b).
\label{eq.4.8}
\end{equation}
As usual, we introduce a cutoff $\omega_{\rm D}$ in calculating the summation of the Matsubara frequency in eq. (\ref{eq.4.8}).
\par
When the interband scattering $v_{ab}$ is absent, ${\bar \omega}^j_m$ and ${\bar {\tilde \Delta}}_j$ satisfy
\begin{equation}
{{\bar \omega}_m^j \over {\bar {\tilde \Delta}}_j}
={\omega_m \over {\tilde \Delta}_j}. 
\label{eq.4.8b}
\end{equation}
In this case, the effect of impurity scatterings in eq. (\ref{eq.4.8}) vanishes as in the case of the nonmagnetic impurity effect in single-band $s$-wave superconductivity.\cite{Maki} As a result, superconductivity is unaffected by impurities. The same situation also occurs when (1) intraband scatterings or (2) interband scatterings are very strong: In the former case, since the first terms in $[\cdot\cdot\cdot]$ in eqs. (\ref{eq.4.4}) and ({\ref{eq.4.5}) are absent in the limit $\alpha_j\to\infty$, eq. (\ref{eq.4.8b}) is satisfied. Physically, when the intraband impurity potential is strong, electrons are excluded from the impurity site. Then the effect of the interband scattering is effectively weakened, so that the situation becomes close to the case of $v_{ab}=0$. In the case (2), since $\eta_j\sim |v_{ab}|^4$ for $|{\bar v}_{ab}|\gg 1$, the first terms in $[\cdot\cdot\cdot]$ in eqs. (\ref{eq.4.4}) and (\ref{eq.4.5}) again vanish in the limit ${\bar v}_{ab}\to\infty$. As discussed in Sec. 3.2, the impurity level becomes $\pm\infty$ in this limit, so that the interband scattering becomes irrelevant. As a result, the situation in the limit $v_{ab}\to \infty$ is equal to the case of $v_{ab}=0$. 
\par
Within the Born approximation, the effect of the interband scattering is controlled by the damping rate $\Gamma_j=\pi n_{\rm imp}N_j(0)|v_{ab}|^2$ ($j=a,b$) when the intraband scattering is absent.\cite{Gusman} Then as shown by Gusman\cite{Gusman} only one averaged energy gap appears in SDOS in the limit $\Gamma_j\to\infty$ ($j=a,b$).\cite{Gusman} (We reproduce this result based in the present model in Appendix C.) Namely, within the Born approximation, the interband scattering simply promotes the mixing of the order parameters. On the other hand, the present results show that this does not always occur; in the limit $v_{ab}\to\infty$, although $\Gamma_j\to\infty$, a two-gap structure corresponding to $|{\tilde \Delta}_a|$ and $|{\tilde \Delta}_b|$ should be observed in SDOS as in the case of $v_{ab}=0$. 
\par
In solving eq. (\ref{eq.4.8}), we employ the following prescription: When we separate eq. (\ref{eq.4.8}) into the real and imaginary parts, we obtain four equations. However, since we can always take $\Delta_a$ real, the number of variables which we have to determine is three, i.e., $\Delta_a$, ${\rm Re}[\Delta_b]$ and ${\rm Im}[\Delta_b]$. Thus we have to eliminate the redundancy in eq. (\ref{eq.4.8}). For this purpose, we rewrite the gap equation as follows: From (\ref{eq.4.4}) and (\ref{eq.4.5}), we find that ${\bar \omega}_m^j$ and ${\bar {\tilde \Delta}}_j$ can be written as
\begin{eqnarray}
\left\{
\begin{array}{l}
\displaystyle
{\rm i}{\bar \omega}_m^j={\rm i}\omega_m{\zeta_j \over F_\omega},
\\
\displaystyle
{\bar {\tilde \Delta}}_j=
{{\bar {\tilde \Delta}}_j' \over F_\omega},
\end{array}
\right.
\label{eq.4.9}
\end{eqnarray}
where ${\bar {\tilde \Delta}}'_j$ $(j=a,b)$ are related to $\Delta_a$ and $\Delta_b$ as 
\begin{eqnarray}
\left\{
\begin{array}{l}
\displaystyle
{\bar {\tilde \Delta}}'_a
=
\zeta_{aa}\Delta_a+{{\bar g}_{ab} \over {\bar g}_{bb}}\zeta_{ab}\Delta_b,\\
\displaystyle
{\bar {\tilde \Delta}}'_b
=
\zeta_{bb}\Delta_b+{{\bar g}_{ba} \over {\bar g}_{aa}}\zeta_{ba}\Delta_a.\\
\end{array}
\right.
\label{eq.4.10}
\end{eqnarray}
In eq. (\ref{eq.4.9}), $F_\omega$ is given by
\begin{equation}
F_\omega=
\bigl(
1-{u_a \over D_a}
{\eta_a \over {\bar \Omega}_a}
\bigr)
\bigl(
1-{u_b \over D_b}
{\eta_b \over {\bar \Omega}_b}
\bigr)
-
{|{\bar v}_{ab}|^4 \over (1+\alpha_a^2)(1+\alpha_b^2)}
{u_au_b \over D_aD_b}
{1 \over {\bar \Omega}_a{\bar \Omega}_b}.
\label{eq.4.11}
\end{equation}
In eqs. (\ref{eq.4.9}) and ({\ref{eq.4.10}), $\zeta_j$ and $\zeta_{jj'}$ $(j,j'=a,b)$ are given by
\begin{eqnarray}
\left\{
\begin{array}{l}
\displaystyle
\zeta_a=
1-
\bigl[
{u_b \over D_b}\eta_b-{u_a \over D_a}
{|{\bar v}_{ab}|^2 \over 1+\alpha_b^2}
\bigr]
{1 \over {\bar \Omega}_b},
\\
\displaystyle
\zeta_b=
1-
\bigl[
{u_a \over D_a}\eta_a-{u_b \over D_b}
{|{\bar v}_{ba}|^2 \over 1+\alpha_a^2}
\bigr]
{1 \over {\bar \Omega}_a},
\end{array}
\right.
\label{eq.4.12}
\end{eqnarray}

\begin{eqnarray}
\left\{
\begin{array}{l}
\displaystyle
\zeta_{aa}=
1-
\bigl[
{u_b \over D_b}\eta_b-
{{\bar g}_{ba} \over {\bar g}_{aa}}
{u_a \over D_a}
{{\bar v}_{ab}^2 \over 1+\alpha_b^2}
\bigr]
{1 \over {\bar \Omega}_b},
\\
\displaystyle
\zeta_{ab}=
1-
\bigl[
{u_b \over D_b}\eta_b-
{{\bar g}_{bb} \over {\bar g}_{ab}}
{u_a \over D_a}
{{\bar v}_{ab}^2 \over 1+\alpha_b^2}
\bigr]
{1 \over {\bar \Omega}_b},
\\
\displaystyle
\zeta_{bb}=
1-
\bigl[
{u_a \over D_a}\eta_a-
{{\bar g}_{ab} \over {\bar g}_{bb}}
{u_b \over D_b}
{{\bar v}_{ba}^2 \over 1+\alpha_a^2}
\bigr]
{1 \over {\bar \Omega}_a},
\\
\displaystyle
\zeta_{ba}=
1-
\bigl[
{u_a \over D_a}\eta_a-
{{\bar g}_{aa} \over {\bar g}_{ba}}
{u_b \over D_b}
{{\bar v}_{ba}^2 \over 1+\alpha_a^2}
\bigr]
{1 \over {\bar \Omega}_a}.
\end{array}
\right.
\label{eq.4.13}
\end{eqnarray}
Substituting eq. (\ref{eq.4.9}) into the gap equation (\ref{eq.4.8}), we obtain
\begin{eqnarray}
\left(
\begin{array}{cc}
\displaystyle
1-{\bar g}_{aa}{\pi \over \beta}\sum_{\omega_m}
s(\omega_m)
{\zeta_{aa} \over {\bar \Omega}'_a} &
\displaystyle
-{\bar g}_{aa}
{{\bar g}_{ab} \over {\bar g}_{bb}}
{\pi \over \beta}\sum_{\omega_m}s(\omega_m){\zeta_{ab}\over {\bar \Omega}'_a} 
\\
\displaystyle
-{\bar g}_{bb}
{{\bar g}_{ba} \over {\bar g}_{aa}}
{\pi \over \beta}\sum_{\omega_m}s(\omega_m){\zeta_{ba}\over {\bar \Omega}'_b} 
&
\displaystyle
1-{\bar g}_{bb}{\pi \over \beta}\sum_{\omega_m}
s(\omega_m)
{\zeta_{bb}\over {\bar \Omega}'_b} 
\end{array}
\right)
\left(
\begin{array}{l}
\Delta_a \\
\Delta_b
\end{array}
\right)
=
\left(
\begin{array}{l}
0 \\
0
\end{array}
\right).
\label{eq.4.14}
\end{eqnarray}
Here ${\bar \Omega}_j'=\sqrt{\zeta_j^2\omega_m^2+|{\bar {\tilde \Delta}}_j'|^2}$ and $s(\omega_m)=F_\omega/|F_\omega|$. Thus, the determinant of the $2\times2$-matrix in eq. (\ref{eq.4.14}) must be zero in order to obtain finite $\Delta_a$ and $\Delta_b$:
\begin{equation}
\bigl(
1-{\bar g}_{aa}{\pi \over \beta}
\sum_{\omega_m}s(\omega_m){\zeta_{aa} \over {\bar \Omega}_a'}
\bigr)
\bigl(
1-{\bar g}_{bb}{\pi \over \beta}
\sum_{\omega_m}s(\omega_m){\zeta_{bb} \over {\bar \Omega}_b'}
\bigr)
-
{\bar g}_{ab}{\bar g}_{ba}
{\pi \over \beta}
\sum_{\omega_m}s(\omega_m){\zeta_{ab} \over {\bar \Omega}_a'}
\cdot
{\pi \over \beta}
\sum_{\omega'_m}s(\omega_m'){\zeta_{ba} \over {\bar \Omega}_b'}=0.
\label{eq.4.15}
\end{equation}
When eq. (\ref{eq.4.15}) is satisfied, the "eigen-value" $(\Delta_a,\Delta_b)$ in eq. (\ref{eq.4.14}) obeys the equation
\begin{equation}
\displaystyle
{\Delta_b \over \Delta_a}=
{
\displaystyle
1-{\bar g}_{aa}
{\pi \over \beta}\sum_{\omega_m}
s(\omega_m)
{\zeta_{aa} \over {\bar \Omega}'_a}
\over 
\displaystyle
{\bar g}_{aa}{{\bar g}_{ab} \over {\bar g}_{bb}}
{\pi \over \beta}\sum_{\omega_m}
s(\omega_m)
{\zeta_{ab} \over {\bar \Omega}_b'}
}.
\label{eq.4.16}
\end{equation}
The real and imaginary parts of Eq. ({\ref{eq.4.16}) give two equations, while eq. (\ref{eq.4.15}) gives only one equation because it is a real equation as proved in Appendix D. Thus we can determine the order parameters $\Delta_a$, ${\rm Re}[\Delta_b]$ and ${\rm Im}[\Delta_b]$ unambiguously from eqs. (\ref{eq.4.15}) and (\ref{eq.4.16}). 
\par
In the clean system, we can take $\zeta_j=\zeta_{jj'}=F_\omega=1$. Then the gap equation (\ref{eq.4.14}) is reduced to eq. (\ref{eq.2.9}) when, as usual, (1) we transform the summations in eq. (\ref{eq.4.14}) to complex integrations $z$, (2) change the variables of the integrations as $z\to\varepsilon\equiv\sqrt{z^2-|{\tilde \Delta}_j|^2}$ and (3) introduce a cutoff $\omega_{\rm D}$.
\par
\vskip3mm
\subsection{Transition temperature}
\vskip2mm
The transition temperature $T_{\rm c}$ is determined by eq. (\ref{eq.4.15}) with $\Delta_a=\Delta_b=0$:
\begin{eqnarray}
0
&=&
\bigl(
1-{\bar g}_{aa}{\pi \over \beta}
\sum_{\omega_m}s^0(\omega_m){\zeta^0_{aa} \over |\zeta^0_a\omega_m|}
\bigr)
\bigl(
1-{\bar g}_{bb}{\pi \over \beta}
\sum_{\omega_m}s^0(\omega_m){\zeta^0_{bb} \over |\zeta^0_b\omega_m|}
\bigr)
\nonumber
\\
&-&
{\bar g}_{ab}{\bar g}_{ba}{\pi \over \beta}
\sum_{\omega_m}s^0(\omega_m){\zeta^0_{ab} \over |\zeta^0_a\omega_m|}\cdot
{\pi \over \beta}
\sum_{\omega'_m}s^0(\omega_m')
{\zeta^0_{ba} \over |\zeta^0_b\omega_m'|}.
\label{eq.4.17}
\end{eqnarray}
Here $\zeta^0_j$ and $\zeta^0_{jj'}$ are given by eqs. (\ref{eq.4.12}) and (\ref{eq.4.13}) with $\Delta_a=\Delta_b=0$:
\begin{eqnarray}
\left\{
\begin{array}{l}
\displaystyle
\zeta^0_a=
1-
\bigl[
{u_b \over D^0_b}\eta_b-{u_a \over D^0_a}
{|{\bar v}_{ab}|^2 \over 1+\alpha_b^2}
\bigr]
{1 \over |\omega_m|+\Gamma_b},
\\
\displaystyle
\zeta^0_b=
1-
\bigl[
{u_a \over D^0_a}\eta_a-{u_b \over D^0_b}
{|{\bar v}_{ba}|^2 \over 1+\alpha_a^2}
\bigr]
{1 \over |\omega_m|+\Gamma_a},
\end{array}
\right.
\label{eq.4.20}
\end{eqnarray}
\begin{eqnarray}
\left\{
\begin{array}{l}
\displaystyle
\zeta^0_{aa}=
1-
\bigl[
{u_b \over D^0_b}\eta_b-
{{\bar g}_{ba} \over {\bar g}_{aa}}
{u_a \over D^0_a}
{{\bar v}_{ab}^2 \over 1+\alpha_b^2}
\bigr]
{1 \over |\omega_m|+\Gamma_b},
\\
\displaystyle
\zeta^0_{ab}=
1-
\bigl[
{u_b \over D^0_b}\eta_b-
{{\bar g}_{bb} \over {\bar g}_{ab}}
{u_a \over D^0_a}
{{\bar v}_{ab}^2 \over 1+\alpha_b^2}
\bigr]
{1 \over |\omega_m|+\Gamma_b},
\\
\displaystyle
\zeta^0_{bb}=
1-
\bigl[
{u_a \over D^0_a}\eta_a-
{{\bar g}_{ab} \over {\bar g}_{bb}}
{u_b \over D^0_b}
{{\bar v}_{ba}^2 \over 1+\alpha_a^2}
\bigr]
{1 \over |\omega_m|+\Gamma_a},
\\
\displaystyle
\zeta^0_{ba}=
1-
\bigl[
{u_a \over D^0_a}\eta_a-
{{\bar g}_{aa} \over {\bar g}_{ba}}
{u_b \over D^0_b}
{{\bar v}_{ba}^2 \over 1+\alpha_a^2}
\bigr]
{1 \over |\omega_m|+\Gamma_a}.
\end{array}
\right.
\label{eq.4.21}
\end{eqnarray}
In eq. (\ref{eq.4.17}), $s^0(\omega_m)=F^0_\omega/|F^0_\omega|$, where $F^0_\omega$ is given by
\begin{equation}
F^0_\omega=
\bigl(
1-{u_a \over D^0_a}
{\eta_a \over |\omega_m|+\Gamma_a}
\bigr)
\bigl(
1-{u_b \over D^0_b}
{\eta_b \over |\omega_m|+\Gamma_b}
\bigr)
-
{|{\bar v}_{ab}|^4 \over (1+\alpha_a^2)(1+\alpha_b^2)}
{u_au_b \over D^0_aD^0_b}
{1 \over (|\omega_m|+\Gamma_a)(|\omega_m|+\Gamma_b)}.
\label{eq.4.22}
\end{equation}
In obtaining eqs. (\ref{eq.4.20})-(\ref{eq.4.22}), we have used the result that eq. (\ref{eq.4.4}) at $T_{\rm c}$ is reduced to
\begin{eqnarray}
\left\{
\begin{array}{l}
\displaystyle
{\rm i}{\bar \omega}^a_m={\rm i}\omega_m
+
{\rm i}
{u_a \over D^0_a}
\bigl[
{|{\bar v}_{ab}|^2 \over 1+\alpha_b^2}
+
\eta_a
\bigr]
{\omega_m \over |\omega_m|}
\equiv
{\rm i}\omega_m+{\rm i}\Gamma_a
{\omega_m \over |\omega_m|}
,
\\
\displaystyle
{\rm i}{\bar \omega}^b_m={\rm i}\omega_m
+
{\rm i}
{u_b \over D^0_b}
\bigl[
{|{\bar v}_{ba}|^2 \over 1+\alpha_a^2}
+
\eta_b
\bigr]
{\omega_m \over |\omega_m|}
\equiv
{\rm i}\omega_m+{\rm i}\Gamma_b
{\omega_m \over |\omega_m|}
,
\end{array}
\right.
\label{eq.4.18}
\end{eqnarray}
where 
\begin{eqnarray}
\left\{
\begin{array}{l}
\displaystyle
D_a^0=
\bigl(
1+{|{\bar v}_{ab}|^2 \over 1+\alpha_b^2}
\bigr)^2
+
\bigl(
\alpha_a-{\alpha_b \over 1+\alpha_b^2}
|{\bar v}_{ab}|^2
\bigr)^2,
\\
\displaystyle
D_b^0=
\bigl(
1+{|{\bar v}_{ba}|^2 \over 1+\alpha_a^2}
\bigr)^2
+
\bigl(
\alpha_b-{\alpha_a \over 1+\alpha_a^2}
|{\bar v}_{ba}|^2
\bigr)^2.
\end{array}
\right.
\label{eq.4.19}
\end{eqnarray}
\par
In the present case, effects of impurity scatterings are controlled by (i) the impurity potential ($v_{aa}$, $v_{bb}$, $v_{ab}$) and (ii) the impurity concentration described by $u_j=n_{\rm imp}/(\pi N_j(0))$. 
In Figs. 3 and 4, we show their effects on $T_{\rm c}$:
\begin{enumerate}
\item[(i)] Figure 3 shows the effects of the interband scattering ${\bar v}_{ab}$ on the transition temperature $T_{\rm c}$. When ${\bar v}_{ab}$ becomes finite, $T_{\rm c}$ decreases with increasing $|{\bar v}_{ab}|$ in the region $0<|{\bar v}_{ab}|\lesssim 1$; however, $T_{\rm c}$ again increases when $|{\bar v}_{ab}|\gesim 1$ reflecting that the interband scattering becomes irrelevant in the limit ${\bar v}_{ab}\to\infty$. When the interband pairing interaction ${\bar g}_{ab}$ is strong, the superconductong states in the two bands strongly couple with each other even in the clean system. In this case, even if transfers of Cooper-pairs occur between the two bands by interband scatterings, this effect would be weak when the phase of the interband scattering $\theta_{ab}$ is absent. Indeed as shown in Fig. 3(a), the suppression of $T_{\rm c}$ by ${\bar v}_{ab}$ is weaker for larger ${\bar g}_{ab}$. However, when the phase of ${\bar v}_{ab}$ is non-zero, the Cooper-pair acquires the phase $2\times\theta_{ab}$ in the transfer between the two bands. Then since the phases of the order parameters in the two bands are the same (${\bar g}_{bb}>0$), the appearance of the Cooper-pair having this extra-phase would destroy superconductivity. Thus $T_{\rm c}$ is more suppressed when $\theta_{ab}$ is finite in Fig. 3(b). On the other hand, since the intraband scattering weakens the effect of the interband scattering, the suppression of $T_{\rm c}$ becomes weak when the intraband scattering becomes finite as shown in Fig. 3(c).
\par
\item[(ii)] Effects of the impurity concentration on $T_{\rm c}$ is shown in Fig. 4. We find that the suppression of $T_{\rm c}$ continues to exist as the impurity concentration increases, which is in contrast to the case of the increase of $|{\bar v}_{ab}|$ in Fig. 3. In the case of $\theta_{ab}=0$ (Fig. 4(a)), the suppression of $T_{\rm c}$ is weaker when the interband pairing ${\bar g}_{ab}$ is stronger. On the other hand, the opposite tendency is obtained in the case of $\theta_{ab}=\pi/2$ as shown in Fig. 4(b). In this former case, when ${\bar g}_{ab}$ is strong, the superconducting states in the two bands strongly coupled with each other as noted in (i). Thus the mixing effect of two order parameters by interband scatterings is weaker for larger ${\bar g}_{ab}$ in the case of $\theta_{ab}=0$. However, in the case of Fig. 4(b) ($\theta_{ab}\ne 0$), since the Cooper-pair acquires the phase $2\theta_{ab}$ in the tunnelling by the interband scattering, the depairing effect becomes more remarkable when ${\bar g}_{ab}$ is stronger. (In the extreme case ${\bar g}_{ab}=0$, one can freely choose the phases of $\Delta_a$ and $\Delta_b$. Then the depairing effect by the tunnelling of the Cooper-pair by the interband scattering can be eliminated by choosing the difference of the phases of $\Delta_a$ and $\Delta_b$ as $2\theta_{ab}$. Indeed, one can eliminate $\theta_{ab}$ from the present model by an appropriate gauge transformation. This means that the depairing effect by $\theta_{ab}$ is absent when ${\bar g}_{ab}=0$.) As a result, in contrast to the case of $\theta_{ab}=0$, the suppression of $T_{\rm c}$ is more remarkable in Fig. 4(b) when the interband pairing ${\bar g}_{ab}$ is stronger.
\end{enumerate}
\vskip3mm
\subsection{Temperature dependence of ${\tilde \Delta}_a$ and ${\tilde \Delta}_b$}
\vskip2mm
Figures 5-7 shows the temperature dependence of ${\tilde \Delta}_a$ and ${\tilde \Delta}_b$.\cite{note3} In Fig. 5, when the impurity concentration ($u_j$) increases, $|{\tilde \Delta}_a|$  and $|{\tilde \Delta}_b|$ approach each other irrespective of $\theta_{ab}$. Although we do not show the result for $0\le |{\bar v}_{ab}|<2$ in Fig. 6, the same tendency as in Fig. 5 is obtained with increasing $|{\bar v}_{ab}|$ up to $|{\bar v}_{ab}|\lesssim 1$. However, when $|{\bar v}_{ab}|\gesim 1$, Fig. 6 shows that difference between $|{\tilde \Delta}_a|$ and $|{\tilde \Delta}_b|$ becomes more remarkable for larger $|{\bar v}_{ab}|$ because of the same reason as that discussed in Secs. 4.2 and 4.3. This tendency is also obtained in Fig. 7 as $\alpha_j$ increases.\par
Figure 5(b) shows that ${\tilde \Delta}_b$ becomes negative when $u$ becomes strong. This is because the Cooper-pair acquires the phase $2\theta_{ab}~(=\pi)$ in the tunnelling by interband scatterings: When this tunnelling occurs frequently, it is favourable that the phases of the order parameters differ by $\pi$ in order to avoid the depairing effect caused by the tunnelling of the Cooper-pairs having the phase $\pi$. This "$\pi$-phase" disappears when the interband scattering becomes very strong: In Fig. 6(b), ${\tilde \Delta}_b$ is positive at ${\bar v}_{ab}=10$. In more detail, the stability of this $\pi$-phase is summarized in Fig. 8: The $\pi$-phase continues to exist when the impurity concentration is high ($u_j\gesim 2$) (Fig. 8(a)), while it disappears when the inter- and/or intra-band scatterings become strong ($|{\bar v}_{ab}|\gesim 3$, $\alpha_j\gesim 1.5$), as shown in panels (b) and (c). 
\vskip3mm
\subsection{Superconducting density of states}
\vskip2mm
Effects of the impurity concentration and the phase/magnitude of the impurity scattering on the mixing of the order parameters have been clarified in Sec. 4.4. In this section, we study how the mixing effect is actually observed in the superconducting density of states (SDOS), because the tunnelling measurement is useful in observing this phenomenon.
\par
The superconducting density of states in the $j$-band is given by
\begin{equation}
N_j(\omega)=
N_j(0)
{\rm Re}
\bigl[
{Z_j(\omega) \over 
\sqrt{Z_j(\omega)^2-{\bar {\tilde \Delta}}^R_j(\omega)^2-{\bar {\tilde \Delta}}^I_j(\omega)^2}
}
\bigr]
=
{\rm Re}
\bigl[
{1 \over 
\sqrt{1-\Xi_j^2}}
\bigr].
\label{eq.4.23}
\end{equation}
Here $Z_j(\omega)={\rm i}{\bar \omega}_m^j({\rm i}\omega_m\to\omega+{\rm i}\delta)$ is the analytic continued renormalized Matsubara frequency in eq. (\ref{eq.4.4}); ${\bar {\tilde \Delta}}^R_j(\omega)$ and ${\bar {\tilde \Delta}}^I_j(\omega)$ are, respectively, the analytic continued quantities of the real and imaginary part of ${\bar {\tilde \Delta}}_j$ in eq. (\ref{eq.4.5}). Their expressions are given by (${\bar {\tilde \Delta}}_j(\omega)^2\equiv {\bar {\tilde \Delta}}^R_j(\omega)^2+{\bar {\tilde \Delta}}^I_j(\omega)^2$)
\begin{eqnarray}
\left\{
\begin{array}{l}
Z_a(\omega)=\omega+{\rm i}{u_a \over D_a(\omega)}
\bigl[
{|{\bar v}_{ab}|^2 \over 1+\alpha_b^2}
{Z_b(\omega) \over \sqrt{Z_b(\omega)^2-{\bar {\tilde \Delta}}_b(\omega)^2}}
+
\eta_a
{Z_a(\omega) \over \sqrt{Z_a(\omega)^2-{\bar {\tilde \Delta}}_a(\omega)^2}}
\bigr],
\\
{\bar {\tilde \Delta}}^R_a(\omega)
={\rm Re}[{\tilde \Delta}_a]
+{\rm i}{u_a \over D_a(\omega)}
\bigl[
{|{\bar v}_{ab}|^2 \over 1+\alpha_b^2}
{
{\bar {\tilde \Delta}}^R_b(\omega)\cos(2\theta_{ab})-{\bar {\tilde \Delta}}^I_b(\omega)\sin(2\theta_{ab})
\over
\sqrt{Z_b(\omega)^2-{\bar {\tilde \Delta}}_b(\omega)^2}
}
+\eta_a
{
{\bar {\tilde \Delta}}_a^R(\omega)
\over
\sqrt{Z_a(\omega)^2-{\bar {\tilde \Delta}}_a(\omega)^2}
}
\bigr],
\\
{\bar {\tilde \Delta}}^I_a(\omega)
={\rm Im}[{\tilde \Delta}_a]
+{\rm i}{u_a \over D_a(\omega)}
\bigl[
{|{\bar v}_{ab}|^2 \over 1+\alpha_b^2}
{
{\bar {\tilde \Delta}}^R_b(\omega)\sin(2\theta_{ab})+{\bar {\tilde \Delta}}^I_b(\omega)\cos(2\theta_{ab})
\over
\sqrt{Z_b(\omega)^2-{\bar {\tilde \Delta}}_b(\omega)^2}
}
+\eta_a
{
{\bar {\tilde \Delta}}_a^I(\omega)
\over
\sqrt{Z_a(\omega)^2-{\bar {\tilde \Delta}}_a(\omega)^2}
}
\bigr],
\end{array}
\right.
\label{eq.4.24}
\end{eqnarray}
where
\begin{equation}
D_a(\omega)=1+\eta_a
+{2|{\bar v}_{ab}|^2 \over 1+\alpha_b^2}
{
Z_a(\omega)Z_b(\omega)-
({\bar {\tilde \Delta}}_a^R{\bar {\tilde \Delta}}_b^R+{\bar {\tilde \Delta}}_a^I{\bar {\tilde \Delta}}_b^I)\cos(2\theta_{ab})-
({\bar {\tilde \Delta}}_b^R{\bar {\tilde \Delta}}_a^I-{\bar {\tilde \Delta}}_b^I{\bar {\tilde \Delta}}_a^R)\sin(2\theta_{ab})
\over
\sqrt{Z_a(\omega)^2-{\bar {\tilde \Delta}}_a(\omega)^2}
\sqrt{Z_b(\omega)^2-{\bar {\tilde \Delta}}_b(\omega)^2}
}.
\label{eq.4.25}
\end{equation}
The valuables in the $b$-band ($Z_b(\omega)$, ${\bar {\tilde \Delta}}_b^R(\omega)$, ${\bar {\tilde \Delta}}_b^I(\omega)$ and $D_b(\omega)$) are also given by eqs. (\ref{eq.4.24}) and (\ref{eq.4.25}) where the band indices '$a$' and '$b$' are interchanged. In calculating SDOS, it is convenient to use the last expression in eq. (\ref{eq.4.23}), where $\Xi_j=\sqrt{{\bar {\tilde \Delta}}^R_j(\omega)^2+{\bar {\tilde \Delta}}^I_j(\omega)^2}/Z_j(\omega)\equiv\sqrt{{\bar \Xi}_j^R(\omega)^2+{\bar \Xi}_j^I(\omega)^2}$:
\begin{eqnarray}
\left\{
\begin{array}{l}
\displaystyle
{\bar \Xi}_a^R
=
{
{\rm Re}[{\tilde \Delta}_a]+{\rm i}
{u_a \over D_a(\omega)}
{|{\bar v}_{ab}|^2 \over 1+\alpha_b^2}
{
{\bar \Xi}_b^R\cos(2\theta_{ab})-{\bar \Xi}_b^I\sin(2\theta_{ab})
\over 
\sqrt{1-\Xi_b^2}
}
\over
\omega+{\rm i}{u_a \over D_a(\omega)}
{|{\bar v}_{ab}|^2 \over 1+\alpha_b^2}
{1 \over \sqrt{1-\Xi_b^2}}
},
\\
\displaystyle
{\bar \Xi}_a^I
=
{
{\rm Im}[{\tilde \Delta}_a]+{\rm i}
{u_a \over D_a(\omega)}
{|{\bar v}_{ab}|^2 \over 1+\alpha_b^2}
{
{\bar \Xi}_b^R\sin(2\theta_{ab})+{\bar \Xi}_b^I\cos(2\theta_{ab})
\over 
\sqrt{1-\Xi_b^2}
}
\over
\omega+{\rm i}{u_a \over D_a(\omega)}
{|{\bar v}_{ab}|^2 \over 1+\alpha_b^2}
{1 \over \sqrt{1-\Xi_b^2}}
}.
\label{eq.4.26}
\end{array}
\right.
\end{eqnarray}
$\Xi_b^R$ and $\Xi_b^I$ are obtained from eq. (\ref{eq.4.26}) by interchanging the indices '$a$' and '$b$'. 
\par
Figures 9 and 10 show impurity effects on SDOS in the cases of $\theta_{ab}=0$ and $\theta_{ab}=\pi/2$, respectively. When the impurity concentration increases, SDOS's in the two bands are mixed with each other to be almost the same at $u=20$ in Fig. 9(a) and $u=5$ in Fig 10(a). In Fig. 10(a), because of the strong depairing effect by the phase of the interband scattering $\theta_{ab}$, the mixed SDOS at $u=5$ has a small energy gap compared with the case of $u=0.1$. This strong depairing effect also leads to a gapless behavior at $u=1$ in Fig. 10(a).
\par
The same mixing effect is observed also in Figs 9(b) and 10(b): Although SDOS's in the two bands are different at $|{\bar v}_{ab}|=0.1$, they become almost the same at $|{\bar v}_{ab}|=1$; however, since the interband impurity effect again becomes weak when $|{\bar v}_{ab}|\gesim 1$, SDOS's differ from each other at $|{\bar v}_{ab}|=100$. Since the effect of the interband scattering is also weakened by the intraband scattering, we obtain two distinct SDOS's at $\alpha=10$ in Figs. 9(c) and 10(c). 
\par
If MgB$_2$ is really a two-band superconductor with two distinct order parameters, we expect the mixing effect in SDOS shown in Figs. 9(a) and 10(a) as the impurity concentration is increased. This effect is independent of the phase of the interband scattering as shown in Figs. 9 and 10 and, in principle, it always occurs when the impurity concentration is high enough. This effect might be useful to judge whether or not superconductivity in MgB$_2$ has two distinct order parameters in the two bands. However, infinitely high impurity concentration is impossible experimentally. Thus when we try to observe the mixing effect, the character of impurity potential is crucial: The intraband scattering must not be strong, and the strength of the interband scattering must be moderate ($|{\bar v}_{ab}|\sim 1$). 
\par
\vskip3mm
\section{Summary}
\vskip2mm
In this paper, we have investigated nonmagnetic impurity effects on two-band $s$-wave superconductivity with two different order parameters. We showed that a bound state appears around a nonmagnetic impurity depending on the character of the interband scattering potential. We also considered the case of finite impurity concentration focusing on the mixing of the two order parameters by interband scatterings. In contrast to the previous works which treat the impurity scattering within the Born approximation, we took into account multi-scattering processes using the $t$-matrix approximation. 
We showed that the mixing of the two order parameters occurs when the impurity concentration becomes high, as discussed previously. However, although the interband scattering is essential for the mixing effect, we clarified that this effect disappears when the interband scattering is very strong ($|{\bar v}_{ab}|\gg 1$.). In addition, the intraband scattering also suppresses the mixing effect. When the magnitude of the impurity potential is very large, a clear two-gap structure is observed in the superconducting density of states as in the case of the clean system (unless the impurity concentration is very high.). 
\par
In order to observe the mixing effect of the order parameters in a two-band superconductor, we have to choose impurities with weak intraband scattering and moderate interband scattering ($|{\bar v}_{ab}|\sim 1$.). If one can choose this kind of impurity for MgB$_2$, it is expected that, if the observed two-gap structure in the superconducting density of state originates from two order parameters in the two bands, the two-gap structure would gradually change into a single-gap structure as the impurity concentration increases. The observation of this impurity effect may be useful in clarifying the origin of the two-gap structure in the superconducting density of states in MgB$_2$.%
%
%
\acknowledgements
The author would like to thank Professor S. Takada for fruitful discussions on the absence of the bound state in the strong interband scattering limit. He also thanks Professor A. Griffin for his hospitality during his stay in University of Toronto. He was financially supported by a Japanese Overseas Research Fellowship.
\newpage
\centerline{\bf Appendices}
\appendix
\section{Equation of Bound State Energy}
From eq. (\ref{eq.3.3}), the energy of the bound state $\omega$ is determined by the equation
\begin{eqnarray}
0&=&{\rm det}
[1-{\hat V}_{aa}({\rm i}\omega_m)\sum_{\sib p}G^0_a({\mib p},{\rm i}\omega_m)]
\nonumber
\\
&=&
{\rm det}\bigl[1-{\hat v}_{ab}{\hat P}_{b}{\hat v}_{ba}{\hat P}_{a}]\bigr]
{\rm det}\bigl[1-{\hat v}_{aa}\sum_{\sib p}G^0_a({\mib p},{\rm i}\omega_m)\bigr],
\label{eq.a.1}
\end{eqnarray}
where ${\hat P}_j=\sum_{\sib p}G^0_j({\mib p},{\rm i}\omega_m)[1-{\hat v}_{jj}\sum_{\sib p}G^0_j({\mib p},{\rm i}\omega_m)]^{-1}$~($j=a,b$). Since ${\hat v}_{aa}$ is the intraband nonmagnetic impurity potential, no pole is obtained from the factor ${\rm det}[1-{\hat v}_{aa}\sum_{\sib p}G^0_a({\mib p},{\rm i}\omega_m)]$ in eq. (\ref{eq.a.1}). Hence we drop this term in what follows. The factor ${\hat P}_j$ can be calculated using  
\begin{equation}
\sum_{\sib p}G^0_j=-\pi N_j(0)
{{\rm i}\omega_m-{\tilde \Delta}_j\rho_1 \over \Omega_j}.
\label{eq.a.2}
\end{equation}
Here $\Omega_j=\sqrt{\omega_m^2+{\tilde \Delta}_j^2}$ and we have taken the order parameters real as ${\hat \Delta}_j={\tilde \Delta}_j\rho_1$. Then we obtain
\begin{equation}
{\hat P}_j=-{\pi N_j(0) \over 1+\alpha_j^2}
\Bigl[
\alpha_j\rho_3+{{\rm i}\omega_m-{\tilde \Delta}_j\rho_1 \over \Omega_j}
\Bigr].
\label{eq.a.3}
\end{equation}
Substituting eqs.(\ref{eq.a.3}) and (\ref{eq.vab}) into eq. (\ref{eq.a.1}), we obtain the equation of the pole as
\begin{eqnarray}
0={\rm det}
\left(
\begin{array}{cc}
1-A & -B \\
B^* & 1-A^*
\end{array}
\right)=
1+|A|^2+|B|^2-(A+A^*),
\label{eq.a.4}
\end{eqnarray}
where the matrix elements $A$ and $b$ are given by
\begin{eqnarray}
\left\{
\begin{array}{l}
\displaystyle
A=
{1 \over (1+\alpha_a^2)(1+\alpha_b^2)}
\left[
|{\bar v}_{ab}|^2
(
{{\rm i}\omega_m \over \Omega_a}+\alpha_a
)
(
{{\rm i}\omega_m \over \Omega_b}+\alpha_b
)
-{\bar v}_{ab}^2
{{\tilde \Delta}_a \over \Omega_a}
{{\tilde \Delta}_b \over \Omega_b}
\right],
\\
\displaystyle
B=
{1 \over (1+\alpha_a^2)(1+\alpha_b^2)}
\left[
|{\bar v}_{ab}|^2
{{\tilde \Delta}_a \over \Omega_a}
(
{{\rm i}\omega_m \over \Omega_b}+\alpha_b
)
+
{\bar v}_{ab}^2
{{\tilde \Delta}_b \over \Omega_b}
(
{{\rm i}\omega_m \over \Omega_a}-\alpha_a
)
\right].
\end{array}
\right.
\label{eq.a.5}
\end{eqnarray}
Calculating each term in eq. (\ref{eq.a.4}) and taking the analytic continuation ${\rm i}\omega_m\to\omega+i\delta$, we obtain eq. (\ref{eq.3.5}). 
\vskip3mm
\section{Derivation of the Renormalized Variables ${\bar \omega}_m^j$ and ${\bar {\tilde \Delta}}_j$}
\vskip3mm
We consider the $a$-band. The self-energy $\Sigma_{aa}$ in eq. (\ref{eq.4.2}) can be rewritten as
\begin{equation}
\Sigma_{aa}({\rm i}\omega_m)=
u_a
{1 \over 
{1 \over \pi N_a(0) {\bar V_{aa}}}-
{1 \over \pi N_a(0)}\sum_{\sib p}G_a({\mib p},{\rm i}\omega_m)
}.
\label{eq.b.1}
\end{equation}
The summation of the Green function in terms of ${\mib p}$ gives
\begin{eqnarray}
{1 \over \pi N_a(0)}\sum_{\sib p}G_a({\mib p},{\rm i}\omega_m)
=
-
{
{\rm i}{\bar \omega}_m^a-{\hat {\bar \Delta}}_a 
\over 
{\bar \Omega}_a
}
&=&
-
{{\rm i}{\bar \omega}_m^a \over {\bar \Omega}_a}
+{|{\bar {\tilde \Delta}}_a| \over {\bar \Omega}_a}
\cos({\bar {\tilde \phi}}_a)\rho_1
-{|{\bar {\tilde \Delta}}_a| \over {\bar \Omega}_a}
\sin({\bar {\tilde \phi}}_a)\rho_2
\nonumber
\\
&\equiv&
\gamma_0+\gamma_1\rho_1+\gamma_2\rho_2,
\label{eq.b.2}
\end{eqnarray}
where ${\bar {\tilde \phi}}_j$ is the phase of ${\bar {\tilde \Delta}}_j$ ($j=a,b$). 
\par
Using the expression of ${\bar V}_{aa}$ in eq. (\ref{eq.4.2}) and eq. (\ref{eq.b.2}), we obtain 
\begin{eqnarray}
\pi N_a(0){\bar V}_{aa}
&=&
-{|{\bar v}_{ab}|^2 \over 1+\alpha_b^2}
{{\rm i}\omega_m \over {\bar \Omega}_b}
-{|{\bar v}_{ab}|^2 \over 1+\alpha_b^2}
{|{\bar {\tilde \Delta}}_b| \over {\bar \Omega}_b}
\cos(2\theta_{ab}+{\bar {\tilde \phi}}_b)\rho_1
\nonumber
\\
&+&
{|{\bar v}_{ab}|^2 \over 1+\alpha_b^2}
{|{\bar {\tilde \Delta}}_b| \over {\bar \Omega}_b}
\sin(2\theta_{ab}+{\bar {\tilde \phi}}_b)\rho_2
+
\bigl(
\alpha_a-{\alpha_b \over 1+\alpha_b^2}|{\bar v}_{ab}|^2
\bigr)\rho_3
\nonumber
\\
&\equiv&
\lambda_0+\lambda_1\rho_1+\lambda_2\rho_2+\lambda_3\rho_3.
\label{eq.b.3}
\end{eqnarray}
Thus
\begin{equation}
{1 \over \pi N_a(0){\bar V}_{aa}}=
{\lambda_1\rho_1+\lambda_2\rho_2+\lambda_3\rho_3-\lambda_0 \over \eta_a},
\label{eq.b.4}
\end{equation}
where $\eta_a=\lambda_1^2+\lambda_2^2+\lambda_3^2-\lambda_0^2$ is reduced to eq. (\ref{eq.4.6}). Substituting eqs. (\ref{eq.b.2}) and (\ref{eq.b.4}) into eq. (\ref{eq.b.1}), we have
\begin{equation}
\Sigma_{aa}({\rm i}\omega_m)=
{u_a \over D_a}
[
(\lambda_0+\eta_a\gamma_0)
+(\lambda_1-\eta_a\gamma_1)\rho_1+(\lambda_2-\eta_a\gamma_2)\rho_2
+\lambda_3\rho_3
].
\label{eq.b.5}
\end{equation}
Here $D_a=[
(\lambda_1-\eta_a\gamma_1)^2+(\lambda_2-\eta_a\gamma_2)^2
+\lambda_3^2-(\lambda_0+\eta_a\gamma_0)^2]/\eta_a$ equals eq. (\ref{eq.4.7}).
Thus from eqs. (\ref{eq.4.1}) and (\ref{eq.b.5}), we find
\begin{eqnarray}
\left\{
\begin{array}{l}
{\rm i}{\bar \omega}_m={\rm i}\omega_m-{u_a \over D_a}(\lambda_0+\eta_a\gamma_0),\\
|{\bar {\tilde \Delta}}_a|\cos({\bar {\tilde \phi}}_a)=
|{\tilde \Delta}_a|\cos({\tilde \phi}_a)-
{u_a \over D_a}(\lambda_1-\eta_a\gamma_1),\\
|{\bar {\tilde \Delta}}_a|\sin({\bar {\tilde \phi}}_a)=
|{\tilde \Delta}_a|\sin({\tilde \phi}_a)+
{u_a \over D_a}(\lambda_2-\eta_a\gamma_2),
\end{array}
\right.
\label{eq.b.6}
\end{eqnarray}
where ${\tilde \phi}_j$ is the phase of ${\tilde \Delta}_j$. Substituting the detailed expressions of $\gamma_i$ and $\lambda_i$ into eq. (\ref{eq.b.6}), we obtain eqs. (\ref{eq.4.4}) and (\ref{eq.4.5}).
\par
We briefly mention that the term proportional to $\rho_3$ in eq. (\ref{eq.b.5}) gives a renormalization of the kinetic energy as 
\begin{equation}
{\tilde \varepsilon}_{\sib p}^a=
\varepsilon_{\sib p}^a+
u_a{\lambda_3 \over D_a}.
\label{eq.b.7}
\end{equation}
This renormalization may be crucial when we consider a correlation function; however, since it does not affect the integration in terms of the kinetic energy within the treatment in this paper, we can safely neglect it.
\par 
\vskip3mm
\section{Mixing of the order parameters within the Born approximation}
\vskip3mm
We consider the case of $\alpha_a=\alpha_b={\bar g}_{ab}={\bar g}_{ba}=0$.\cite{Gusman} In this case, since the phase of the interband scattering $\theta_{ab}$ is irrelevant, we can take $v_{ab}$ real. In addition, ${\tilde \Delta}_j=\Delta_j$ and $\Delta_b$ can be also taken real. When we retain scattering processes within the Born approximation, eqs. (\ref{eq.4.4}) and (\ref{eq.4.5}) are reduced to
\begin{eqnarray}
\left\{
\begin{array}{l}
\displaystyle
{\rm i}{\bar \omega}^a_m={\rm i}\omega_m
+\Gamma_a
{{\rm i}{\bar \omega}_m^b \over {\bar \Omega}_b},
\\
\displaystyle
{\rm i}{\bar \omega}^b_m={\rm i}\omega_m
+\Gamma_b
{{\rm i}{\bar \omega}_m^a \over {\bar \Omega}_a},
\end{array}
\right.
\label{eq.d.1}
\end{eqnarray}
\begin{eqnarray}
\left\{
\begin{array}{l}
\displaystyle
{\bar {\tilde \Delta}}_a=\Delta_a
+\Gamma_a
{{\bar {\tilde \Delta}}_b \over {\bar \Omega}_b},
\\
\displaystyle
{\bar {\tilde \Delta}}_b=\Delta_b
+\Gamma_b
{{\bar {\tilde \Delta}}_a \over {\bar \Omega}_a},
\end{array}
\right.
\label{eq.d.2}
\end{eqnarray}
where $\Gamma_a$ and $\Gamma_b$ are defined in eq. (\ref{eq.4.18}) where we take $\alpha_a=\alpha_b=\eta_a=\eta_b=0$ and $D_a^0=D_b^0=1$. Equations (\ref{eq.d.1}) and (\ref{eq.d.2}) can be rewritten as
\begin{eqnarray}
\left\{
\begin{array}{l}
\displaystyle
{\bar \omega}_r^a=1
+\Gamma_a
{{\bar \omega}_r^b \over {\bar \Omega}_b},
\\
\displaystyle
{\bar \omega}^b_r=1
+\Gamma_b
{{\bar \omega}_r^a \over {\bar \Omega}_a},
\end{array}
\right.
\label{eq.d.3}
\end{eqnarray}
\begin{eqnarray}
\left\{
\begin{array}{l}
\displaystyle
{\bar \Delta}_r^a=1
+X\Gamma_a
{{\bar \Delta}_r^b \over {\bar \Omega}_b},
\\
\displaystyle
{\tilde \Delta}_r^b=1
+{\Gamma_b \over X}
{{\bar \Delta}_r^a \over {\bar \Omega}_a}.
\end{array}
\right.
\label{eq.d.4}
\end{eqnarray}
Here we have introduced ${\bar \omega}_r^j\equiv{\bar \omega}_m^j/\omega_m$, ${\bar \Delta}_r^j\equiv{\bar {\tilde \Delta}}_j/\Delta_j$ and $X\equiv\Delta_b/\Delta_a$. When we define $Y_j\equiv {\bar \Delta}_r^j/{\bar \omega}_r^j$, we find a relation $Y_a=XY_b$ in the dirty limit ($\Gamma_a\gg 1$ and $\Gamma_b\gg 1$). Using this relation, we find that $Y_a$ and $Y_b$ obey the following equation:
\begin{eqnarray}
\left(
\begin{array}{cc}
1+{\Gamma_a \over \Omega_r} & -{\Gamma_a \over X\Omega_r} \\
-{\Gamma_b \over X\Omega_r} & 1+{\Gamma_b \over \Omega_r} 
\end{array}
\right)
\left(
\begin{array}{c}
Y_a \\ Y_b
\end{array}
\right)
=
\left(
\begin{array}{c}
1 \\ 1
\end{array}
\right),
\label{eq.d.5}
\end{eqnarray}
where $\Omega_r=\sqrt{\omega_m^2+Y_b^2{\tilde \Delta}_b^2}$. The solution of eq. (\ref{eq.d.5}) is given by $Y_j={\Delta}_{AV}/\Delta_j$ (in the dirty limit), where
\begin{equation}
{\Delta}_{AV}=
{{\Delta}_aN_a(0)+{\Delta}_bN_b(0)
\over 
N_a(0)+N_b(0)}
\label{eq.d.6}
\end{equation}
is an averaged order parameter. Then the gap equation (\ref{eq.4.8}) is reduced to
\begin{equation}
1={\bar g}_{jj}Y_j{\pi \over \beta}
\sum_{\omega_m}
{1 \over \sqrt{\omega_m^2+\Delta_{AV}^2}}.
\label{eq.d.7}
\end{equation}
Substituting the explicit form of $Y_j$ into eq. (\ref{eq.d.7}) we obtain
\begin{eqnarray}
\left(
\begin{array}{cc}
\displaystyle
1-{{\bar g}_{aa}N_a(0) \over N_a(0)+N_b(0)}
{\pi \over \beta}\sum_{\omega_m}
{1 \over \sqrt{\omega_m^2+\Delta_{AV}^2}}
&
\displaystyle
-{{\bar g}_{aa}N_b(0) \over N_a(0)+N_b(0)}
{\pi \over \beta}\sum_{\omega_m}
{1 \over \sqrt{\omega_m^2+\Delta_{AV}^2}}
\\
\displaystyle
-{{\bar g}_{bb}N_a(0) \over N_a(0)+N_b(0)}
{\pi \over \beta}\sum_{\omega_m}
{1 \over \sqrt{\omega_m^2+\Delta_{AV}^2}}
&
\displaystyle
1-{{\bar g}_{bb}N_b(0) \over N_a(0)+N_b(0)}
{\pi \over \beta}\sum_{\omega_m}
{1 \over \sqrt{\omega_m^2+\Delta_{AV}^2}}
\end{array}
\right)
\left(
\begin{array}{c}
\Delta_a \\ \Delta_b
\end{array}
\right)
=0.
\nonumber
\\
\label{eq.d.8}
\end{eqnarray}
Equation (\ref{eq.d.8}) has a solution when the following equation is satisfied:
\begin{equation}
1=
{{\bar g}_{aa}N_a(0)+{\bar g}_{bb}N_b(0) \over N_a(0)+N_b(0)}
{\pi \over \beta}\sum_{\omega_m}
{1 \over \sqrt{\omega_m^2+\Delta_{AV}^2}}.
\label{eq.d.9}
\end{equation}
Equation (\ref{eq.d.9}) can be interpreted as the gap equation of $\Delta_{AV}$ with the averaged pairing interaction $({\bar g}_{aa}N_a(0)+{\bar g}_{bb}N_b(0))/(N_a(0)+N_b(0))$. 
\par
In the dirty limit, we obtain the following relation:
\begin{equation}
{
{\bar \omega}_m^j
\over
\sqrt{{\bar \omega}_m^j{}^2+{\bar {\tilde \Delta}}_j^2}
}
=
{\omega_m \over 
\sqrt{\omega_m^2+\Delta_{AV}^2}}~~~~(\omega_m>0).
\label{eq.d.10}
\end{equation}
Since the superconducting density of states in the $j$-band is obtained from the analytic continuation of LHS in eq. (\ref{eq.d.10}), we find that the excitation gap in both bands is equal to $\Delta_{AV}$. Namely, in the dirty limit, only one excitation gap $\Delta_{AV}$ appears in the superconducting density of states within the Born approximation.
\vskip3mm
\section{Evaluation of Eq. (\ref{eq.4.15})}
\vskip2mm
When we expand eq. (\ref{eq.4.15}), we obtain
\begin{eqnarray}
1
&=&
{\bar g}_{aa}{\pi \over \beta}
\sum_{\omega_m}s(\omega_m){\zeta_{aa} \over {\bar \Omega}_a'}
+
{\bar g}_{bb}{\pi \over \beta}
\sum_{\omega_m}s(\omega_m){\zeta_{bb} \over {\bar \Omega}_b'}
\nonumber
\\
&-&
{\bar g}_{aa}{\bar g}_{bb}
{\pi \over \beta}
\sum_{\omega_m}s(\omega_m){\zeta_{aa} \over {\bar \Omega}_a'}
\cdot
{\pi \over \beta}
\sum_{\omega_m'}s(\omega_m'){\zeta_{bb} \over {\bar \Omega}_b'}
-
{\bar g}_{ab}{\bar g}_{ba}{\pi \over \beta}
\sum_{\omega_m}s(\omega_m){\zeta_{ab} \over {\bar \Omega}_a'}
\cdot
{\pi \over \beta}
\sum_{\omega_m'}s(\omega_m'){\zeta_{ba} \over {\bar \Omega}_b'}.
\label{eq.c.1}
\end{eqnarray}
The first and the second terms in RHS in eq. (\ref{eq.c.1}) include the following complex terms:
\begin{equation}
{\pi \over \beta}
\sum_{\omega_m}
s(\omega_m)
\bigl[
{\bar g}_{ba}{u_a \over D_a}{{\bar v}_{ab}^2 \over 1+\alpha_b^2}
{1 \over {\bar \Omega}_b{\bar \Omega}_a'}
+
{\bar g}_{ab}{u_b \over D_b}{{\bar v}_{ba}^2 \over 1+\alpha_a^2}
{1 \over {\bar \Omega}_a{\bar \Omega}_b'}
\bigr].
\label{eq.c.2}
\end{equation}
However, from eqs. (\ref{eq.4.7}) and (\ref{eq.4.9}) and the definition of $u_j$, we obtain $(1+\alpha_b^2)D_a=(1+\alpha_a^2)D_b$ and $N_a(0)u_a=N_b(0)u_b$. Substituting these relations into eq. (\ref{eq.c.2}) and noting that $g_{ab}=g_{ba}$ and ${\bar \Omega}_j={\bar \Omega}_j'/|F_{\omega}|$, we find that eq. (\ref{eq.c.2}) is real. Thus the first and the second terms in eq. (\ref{eq.c.1}) is also real. 
\par
The third and the fourth terms in eq. (\ref{eq.c.1}) can be rewritten as
\begin{eqnarray}
({\bar g}_{aa}{\bar g}_{bb}&-&{\bar g}_{ab}{\bar g}_{ba})
{\pi^2 \over \beta^2}\sum_{\omega_m\omega_m'}
{
s(\omega_m)s(\omega_m')
\over {\bar \Omega}'_a(\omega_m){\bar \Omega}'_b(\omega_m')}
\bigl[
\bigl(
1-{u_b \over D_b(\omega_m)}
{\eta_b \over {\bar \Omega}_b(\omega_m)}
\bigr)
\bigl(
1-{u_a \over D_a(\omega_m')}
{\eta_a \over {\bar \Omega}_a(\omega_m')}
\bigr)
\nonumber
\\
&-&
{u_au_b|{\bar v}_{ab}|^4 \over 
(1+\alpha_a^2)(1+\alpha_b^2)D_a(\omega_m)D_b(\omega_m')
{\bar \Omega}_a(\omega_m){\bar \Omega}_b(\omega_m')
}
\bigr].
\label{eq.c.3}
\end{eqnarray}
Equation (\ref{eq.c.3}) is real, so that eq. (\ref{eq.4.15}) is found real.
\vskip5mm

\newpage

\newpage
\centerline{\bf Figure Captions}
\begin{enumerate}
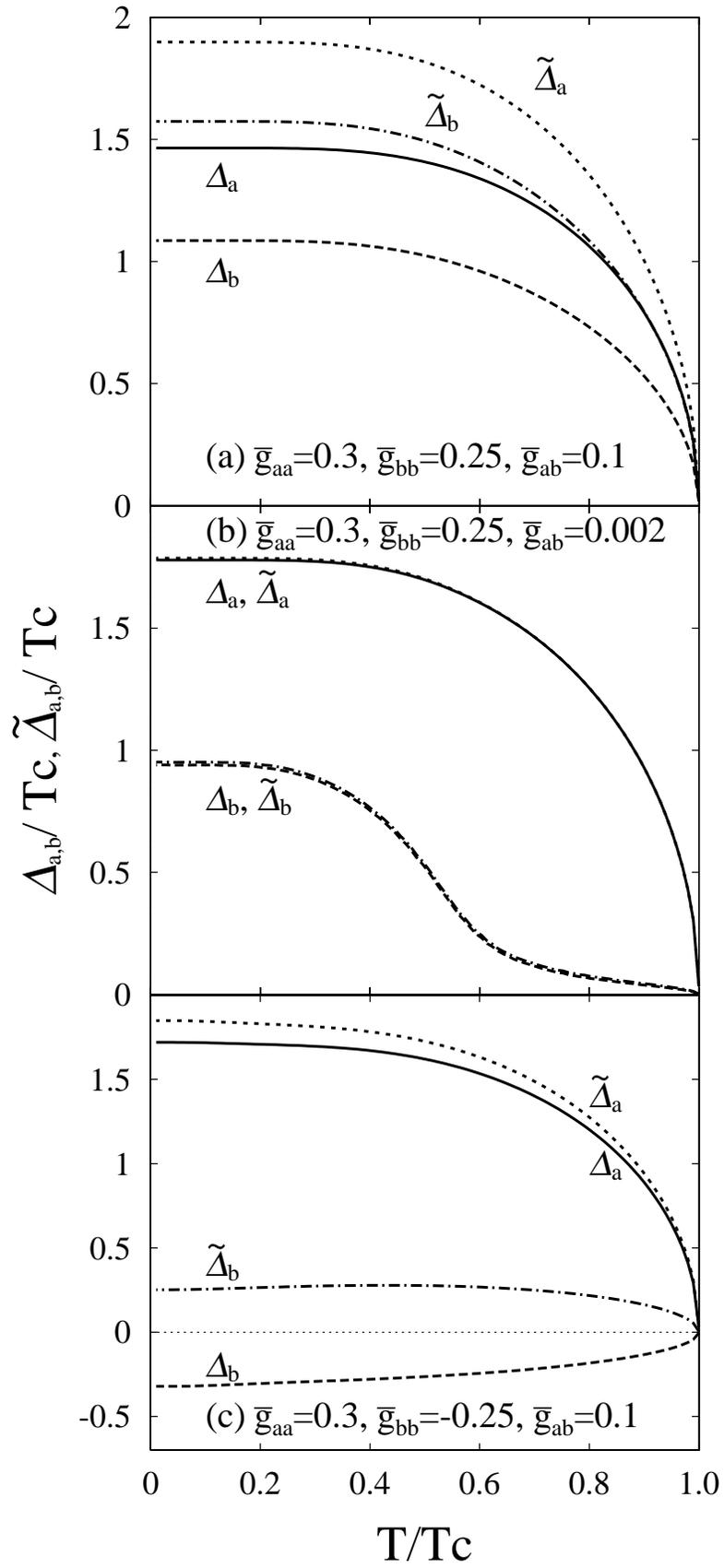
\item[Fig.1:] Temperature dependence of the order parameters in the clean system. We put $N_a(0)=N_b(0)$. This condition is used throughout this paper. In this case, ${\bar g}_{ab}={\bar g}_{ba}$ is satisfied. In panels (a) and (b), the intraband interactions ${\bar g}_{aa}$ and ${\bar g}_{bb}$ are attractive; panel (a) shows the case of ${\bar g}_{ab}\sim ({\bar g}_{aa},{\bar g}_{bb})$, while panel (b) is the case of ${\bar g}_{ab}\ll({\bar g}_{aa},{\bar g}_{bb})$. Panel (c) shows the case when ${\bar g}_{bb}$ is repulsive.
\par

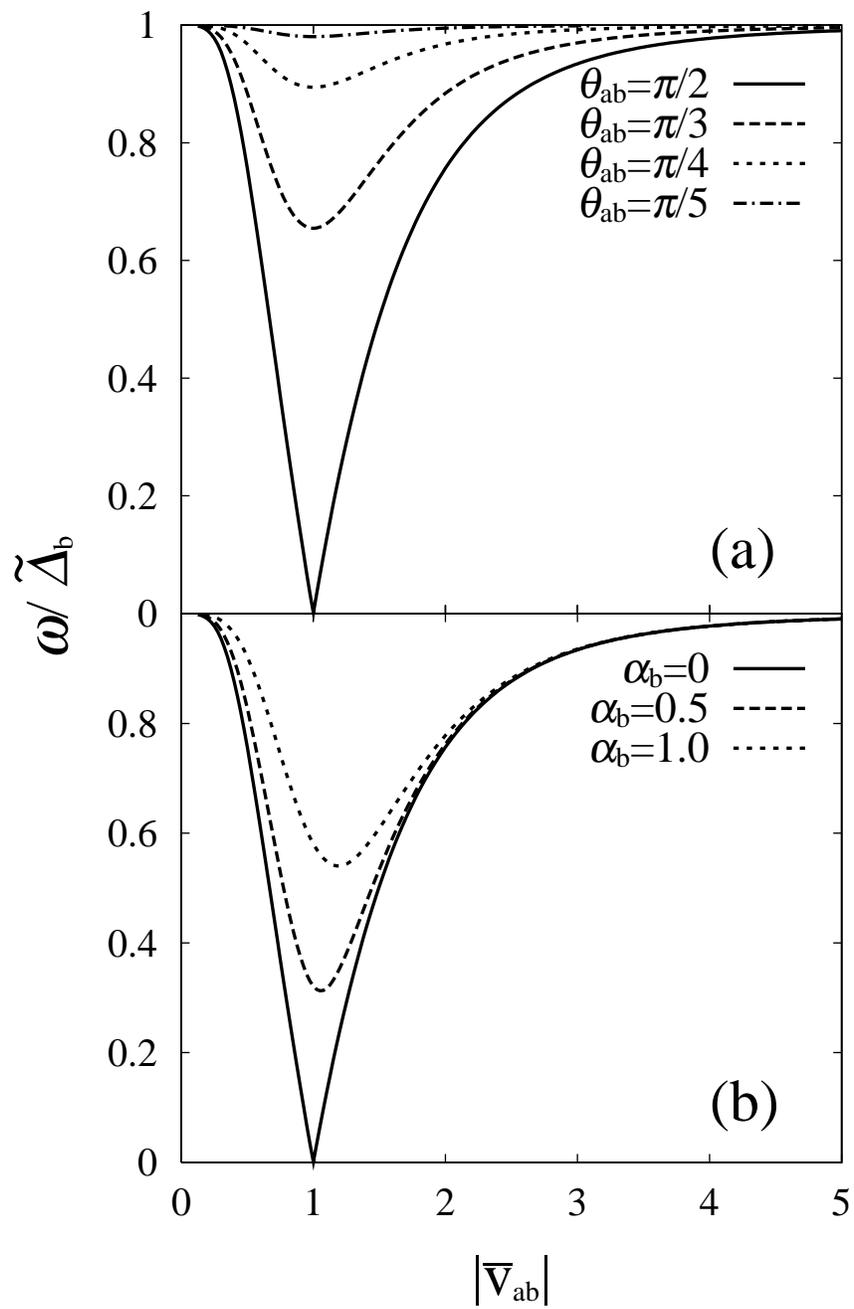
\item[Fig.2:] Dependence of the bound state energy on the interband impurity scattering ${\bar v}_{ab}$. We take ${\tilde \Delta}_a=2$, ${\tilde \Delta}_b=1$ and $\alpha_a=0$. Panel (a) shows the effect of the phase of ${\bar v}_{ab}$ in the case of $\alpha_b=0$. Since ${\tilde \Delta}_b^2-\cos(2\theta_{ab}){\tilde \Delta}_a{\tilde \Delta}_b\ge 0$ for $\theta_{ab}\ge\pi/6$, the bound state is absent for $\theta_{ab}\ge\pi/6$. Panel (b) shows the effect of the intraband potential scattering $\alpha_b$ in the $b$-band in the case of $\theta_{ab}=\pi/2$.
\par

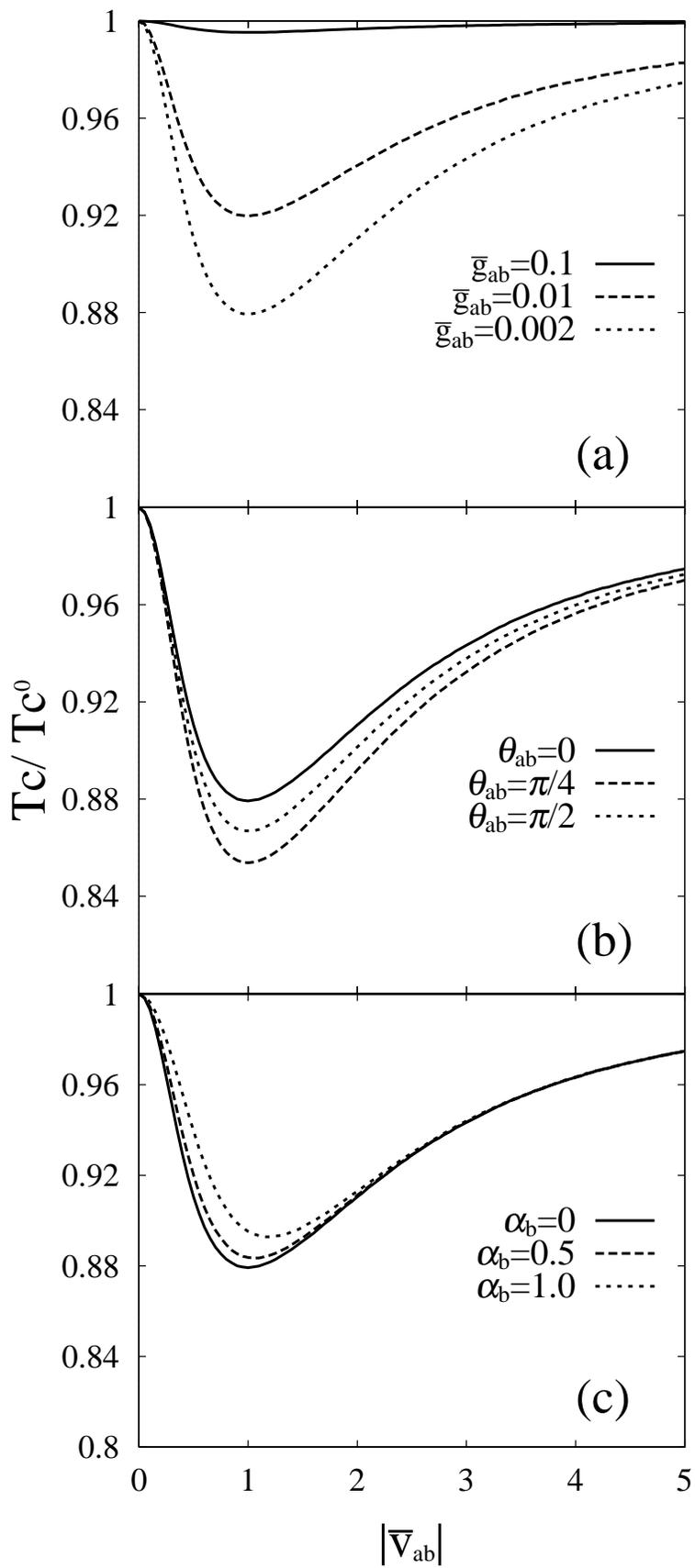
\item[Fig.3:] Dependence of the transition temperature $T_{\rm c}$ on the interband potential scattering ${\bar v}_{ab}$. The transition temperature is normalized by the value at ${\bar v}_{ab}=0$. We take ${\bar g}_{aa}=0.3$, ${\bar g}_{bb}=0.25$, $u_a=u_b=1$ and $\alpha_a=0$. (a) Effect of the interband pairing interaction ${\bar g}_{ab}$ in the case of $\theta_{ab}=0$ and $\alpha_b=0$. (b) Effect of the phase of the interband potential scattering $\theta_{ab}$ in the case of ${\bar g}_{ab}=0.002$ and $\alpha_b=0$. (b) Effect of the intraband potential scattering in the $b$-band ($\alpha_b$). We put ${\bar g}_{ab}=0.002$ and $\theta_{ab}=0$.
\par

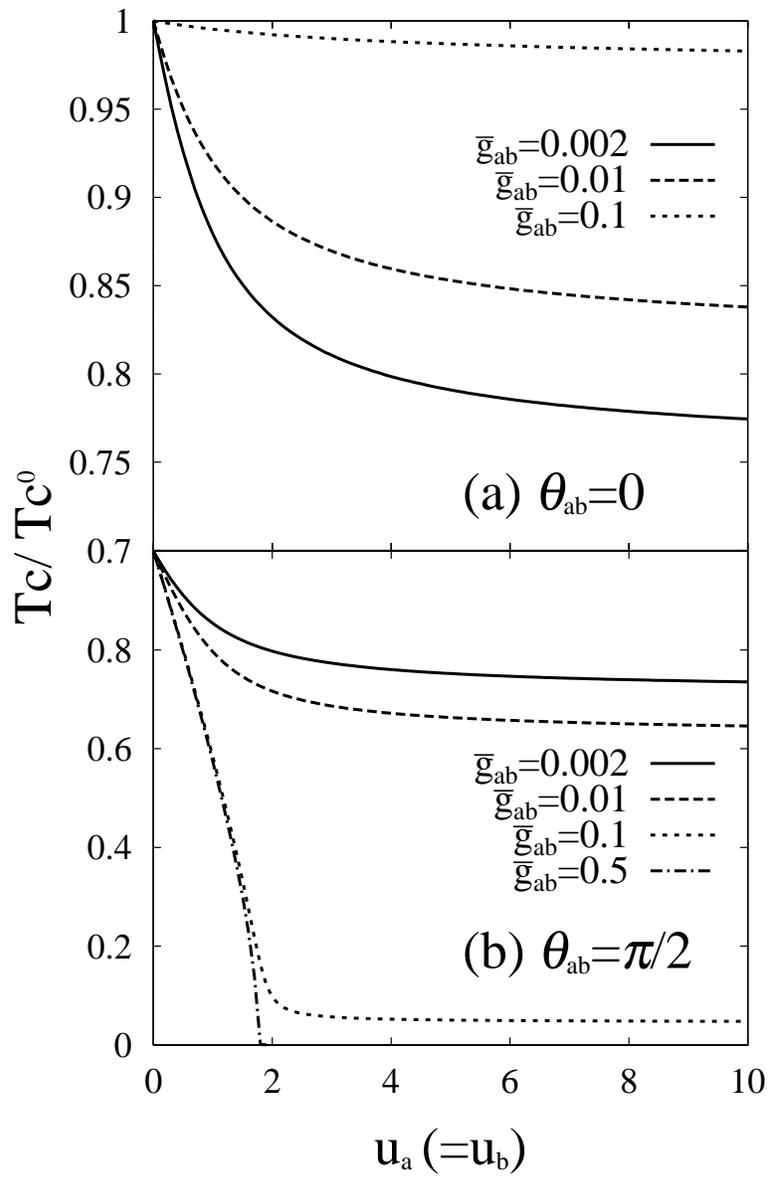
\item[Fig.4:] Dependence of the transition temperature $T_{\rm c}$ on the impurity concentration $u_j=n_{\rm imp}/(\pi N_j(0))$. In this figure, we take $u_a=u_b$ and $T_{\rm c}$ is normalized by the value in clean system ($T_{\rm c}^0$). The other parameters are taken as ${\bar g}_{aa}=0.3$, ${\bar g}_{bb}=0.25$, $\alpha_a=\alpha_b=0$ and $|{\bar v}_{ab}|=1$. In panel (a), the phase of the interband scattering is taken $\theta_{ab}=0$, while in panel (b) $\theta_{ab}=\pi/2$.
\par

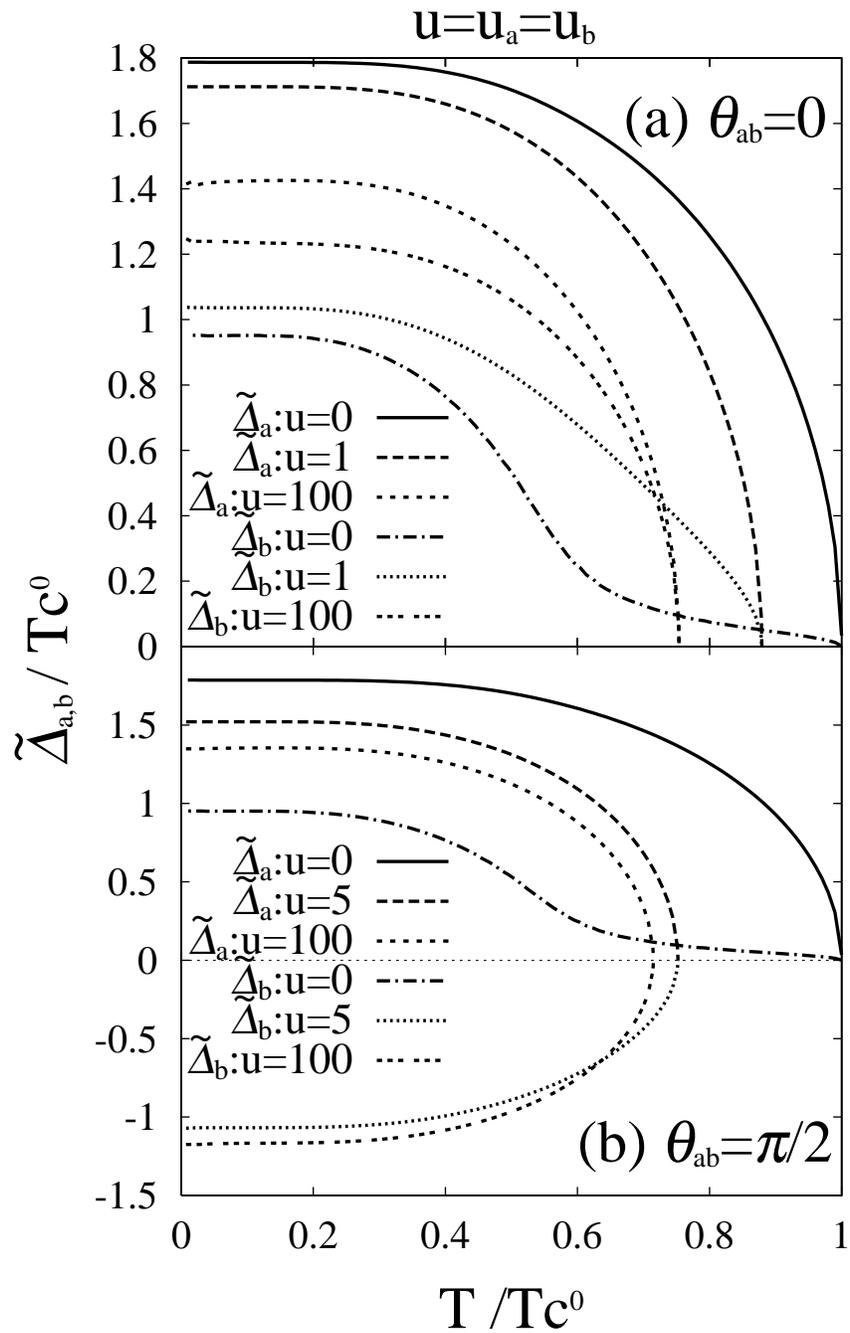
\item[Fig.5:] Temperature dependence of ${\tilde \Delta}_a$ and ${\tilde \Delta}_b$ in the cases of (a) $\theta_{ab}=0$ and (b) $\theta_{ab}=\pi/2$. We take ${\bar g}_{aa}=0.3$, ${\bar g}_{bb}=0.25$, ${\bar g}_{ab}=0.002$, $\alpha_a=\alpha_b=0$ and $|{\bar v}_{ab}|=1$. The $x$- and $y$-axis are normalized by $T_{\rm c}$ in the clean system.
\par

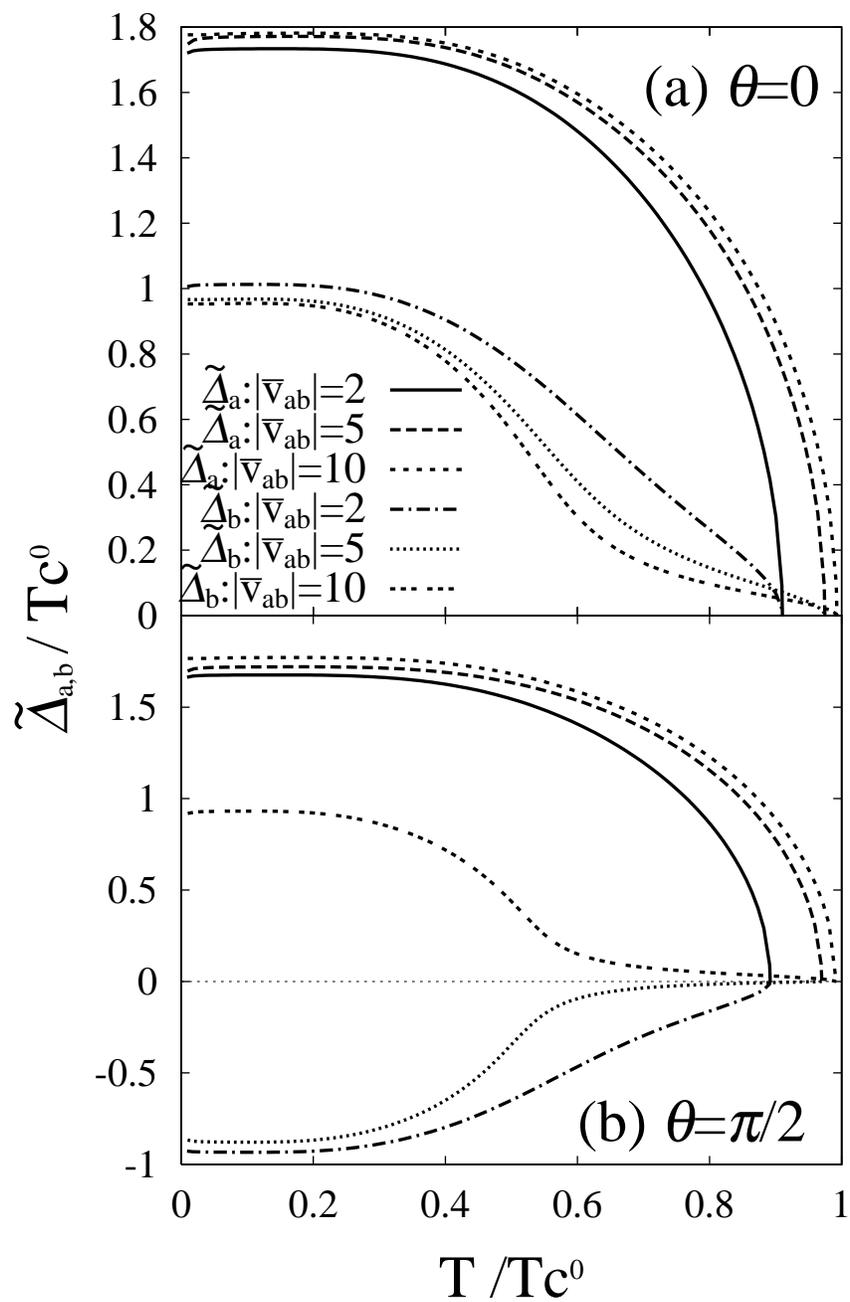
\item[Fig.6:] Effect of the interband scattering on temperature dependence of ${\tilde \Delta}_a$ and ${\tilde \Delta}_b$ in the cases of (a) $\theta_{ab}=0$ and (b) $\theta_{ab}=\pi/2$. We take ${\bar g}_{aa}=0.3$, ${\bar g}_{bb}=0.25$, ${\bar g}_{ab}=0.002$, $\alpha_a=\alpha_b=0$ and $u_a=u_b=1$.
\par

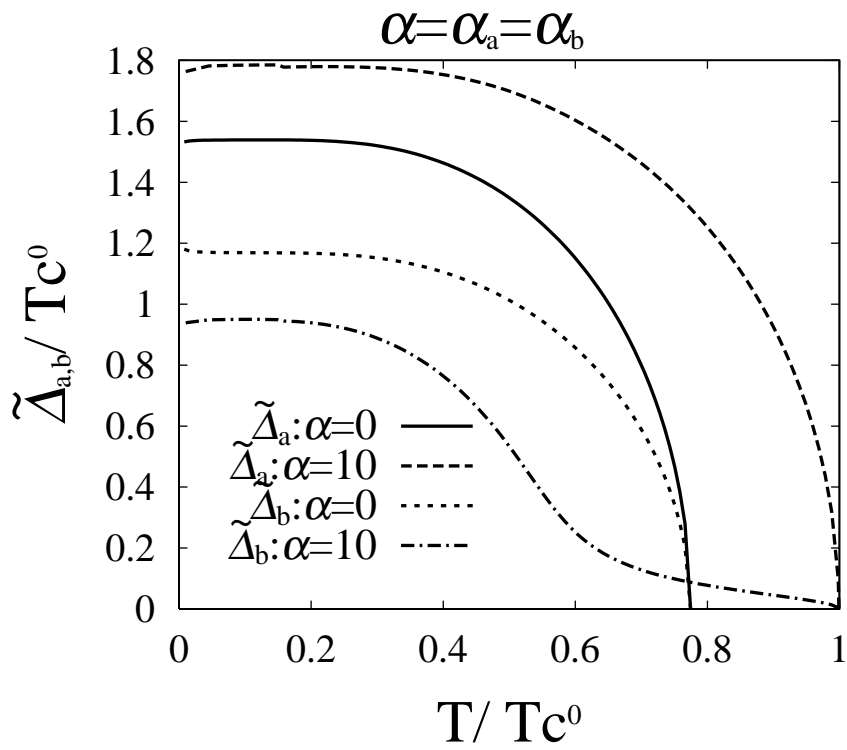
\item[Fig.7:] Effect of intraband scattering ($\alpha_a$ and $\alpha_b$) on $T_{\rm c}$. The figure shows the case of $\theta_{ab}=0$ (${\bar v}_{ab}=1$). The other parameters are ${\bar g}_{aa}=0.3$, ${\bar g}_{bb}=0.25$, ${\bar g}_{ab}=0.002$, $u_a=u_b=10$ and $|{\bar v}_{ab}|=1$.

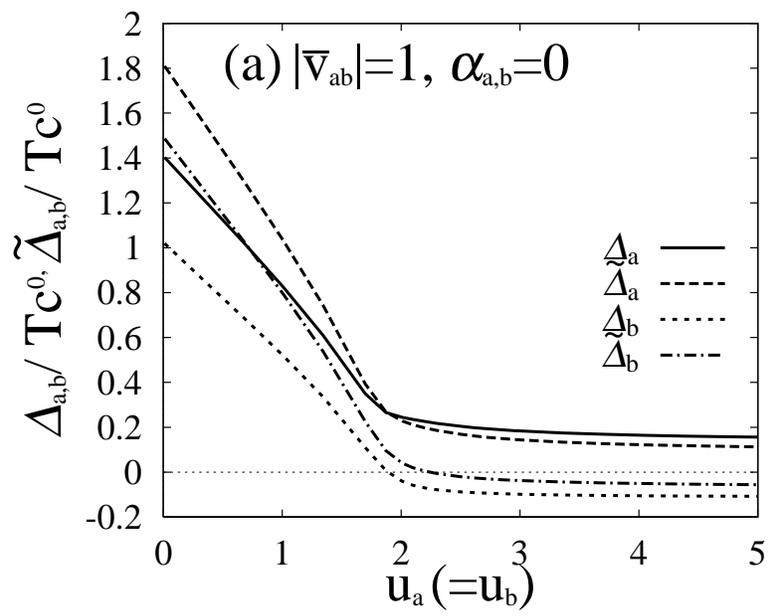
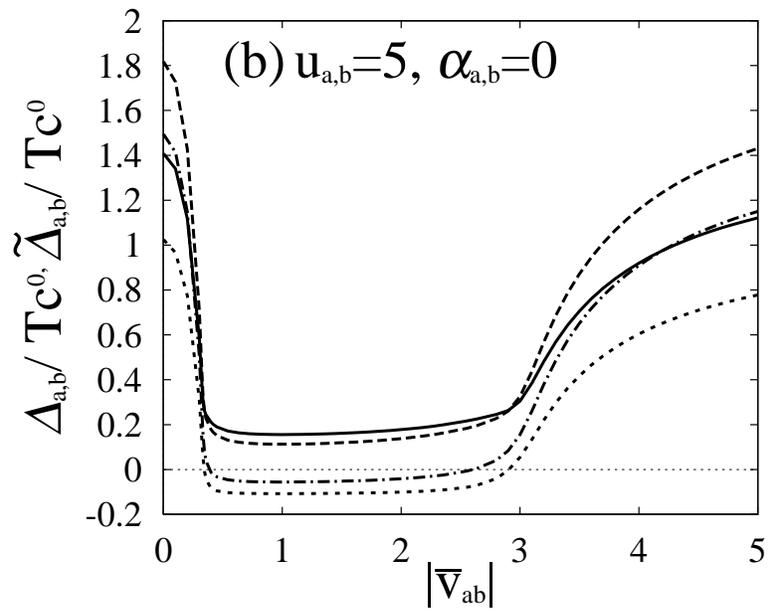
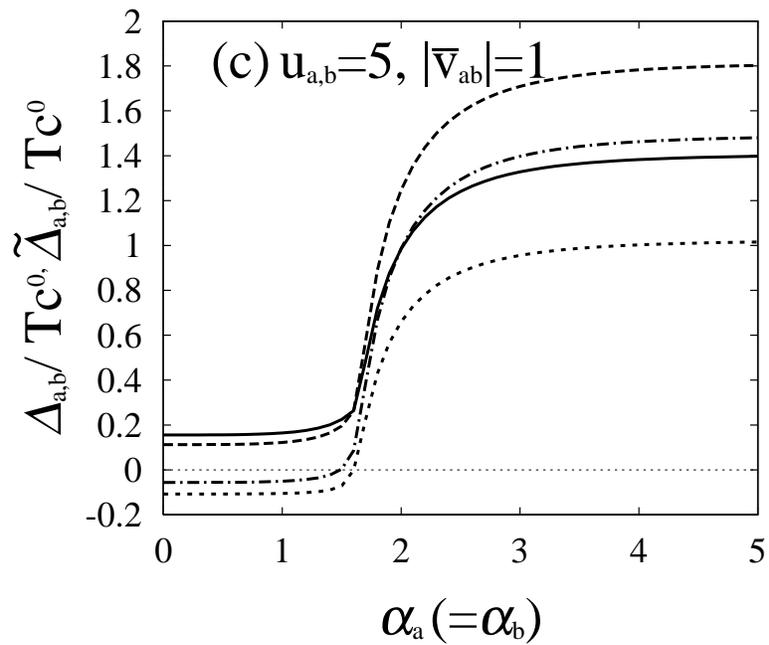
\item[Fig.8:] Stability of "$\pi$-phase" caused by the interband scattering with $\theta_{ab}=\pi/2$. We take $T=0.5T_{\rm c}$, ${\bar g}_{aa}=0.3$, ${\bar g}_{bb}=0.25$ and ${\bar g}_{ab}=0.1$.

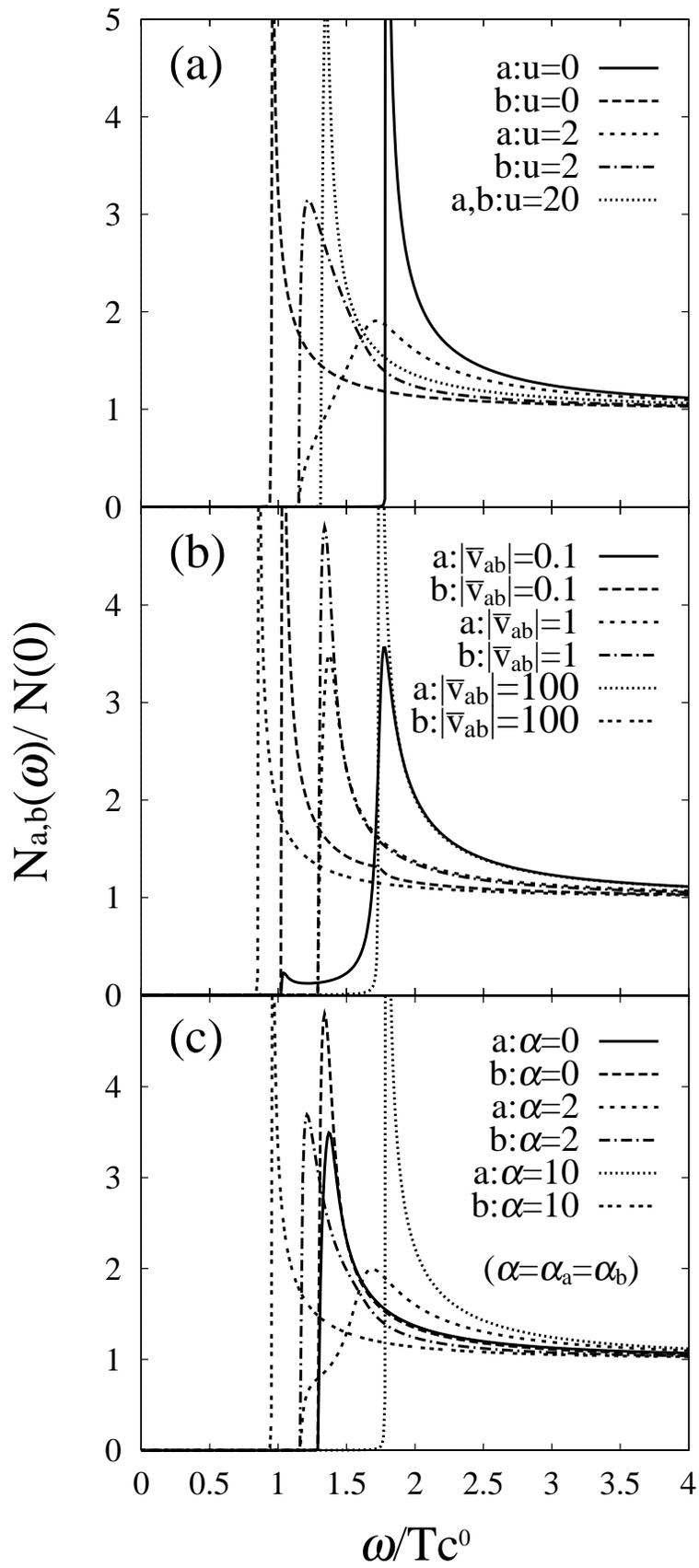
\item[Fig.9:] Superconducting density of states at $T=0.1T_{\rm c}$ in the case of $\theta_{ab}=0$. We take ${\bar g}_{aa}=0.3$, ${\bar g}_{bb}=0.25$ and ${\bar g}_{ab}=0.002$. $N_j(\omega)$ is normalized by the density of states at the Fermi level in the normal state ($N_a(0)=N_b(0)\equiv N(0)$). In this figure, 'a' and 'b' mean SDOS in the $a$- and $b$-band, respectively. (a) Effect of impurity concentration ($u=u_a=u_b$) at $\alpha_a=\alpha_b=0$ and $|{\bar v}_{ab}|=1$. Since SDOS's are almost the same in the two bands at $u=5$, we show one of them only in the figure. (b) Effect of the magnitude of the interband scattering $|{\bar v}_{ab}|$ at $\alpha_a=\alpha_b=0$ and $u_a=u_b=10$. (b) Effect of the intraband scattering $\alpha~(=\alpha_a=\alpha_b)$  at $|{\bar v}_{ab}|=1$ and $u_a=u_b=10$. 
\par

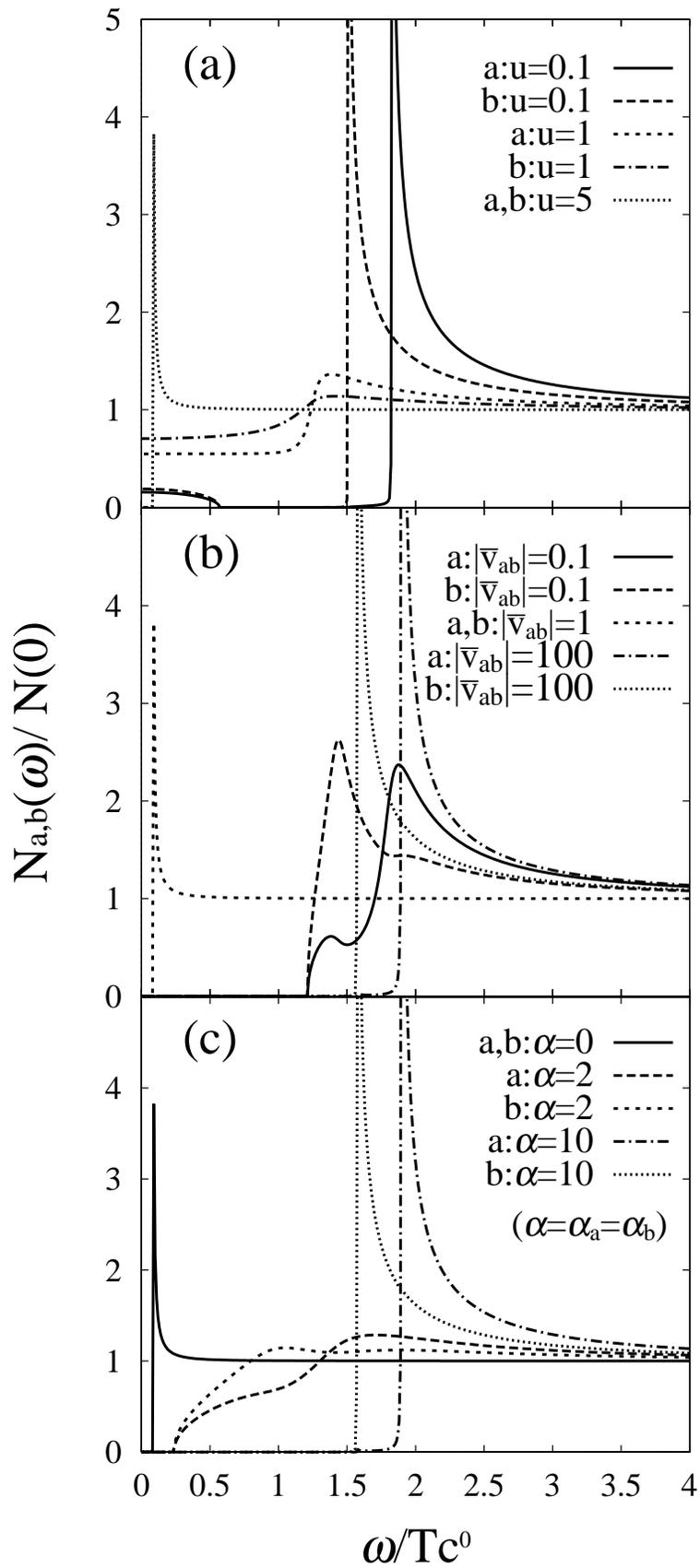
\item[Fig.10:] Superconducting density of states at $T=0.1T_{\rm c}$ in the case of $\theta_{ab}=\pi/2$. We take ${\bar g}_{aa}=0.3$, ${\bar g}_{bb}=0.25$ and ${\bar g}_{ab}=0.1$. (a) Effect of $u~(=u_a=u_b)$ at $\alpha_a=\alpha_b=0$ and $|{\bar v}_{ab}|=1$. (b) Effect of $|{\bar v}_{ab}|$ at $\alpha_a=\alpha_b=0$ and $u_a=u_b=5$. (c) Effect of the intraband scattering $\alpha$ at $|{\bar v}_{ab}|=1$ and $u_a=u_b=5$. Since $N_a(\omega)$ and $N_b(\omega)$ are almost the same at $u=5$ in panel (a), $|{\bar v}_{ab}|=100$ in panel (b) and  $\alpha=0$ in panel (c), only $N_a(\omega)$ is shown in these cases. 
\end{enumerate}

\begin{thebibliography}{99}
\bibitem{Akimitsu} J. Nagamatsu, N. Nakagawa, T. Muranaka, Y. Zenitani and J. Akimitsu: Nature {\bf 410} (2001) 63.
\bibitem{Bou} F. Bouquet, R. A. Fisher, N. E. Phillips, D. G. Hinks and J. D. Jorgensen: Phys. Rev. Lett. {\bf 87} (2001) 047001.
\bibitem{Man} F. Manzano and A. Carrington: cond-mat/0110109.
\bibitem{Bud} S. L. Bud'ko, G. Lapershot, C. Petrovic, C. E. Cummingham, N. Anderson and P. C. Canfield: Phys. Rev. Lett. {\bf 86} (2001) 1877.
\bibitem{Yamaji} K. Yamaji: J. Phys. Soc. Jpn. {\bf 70} (2001) 1476.
\bibitem{Furukawa} N. Furukawa: J. Phys. Soc. Jpn. {\bf 70} (2001) 1483.
\bibitem{Imada} M. Imada: cond-mat/0103006.
\bibitem{Shulga} S. V. Shulga, S.-L. Drechsler, H. Echrig, H. Rosner and W. Pickett: cond-mat/0103154.
\bibitem{Liu} A. Y. Liu, I. I. Mazin and J. Kortus: Phys. Rev. Lett. {\bf 87} (2001) 087005.
\bibitem{Golubov} A. A. Golubov, J. Kortus, O. V. Dolgov, O. Jepsen, Y. Kong, O. K. Andersen, B. J. Gibson, K. Ahn and R. K. Kremer: cond-mat/0111262.
\bibitem{Nakai} N. Nakai, M. Ichioka and K. Machida: cond-mat/0111088.
\bibitem{Giu} F. Giubileo, D. Roditchev, W. Sacks, R. Lamy, D. X. Thanh, J. Klein, S. Miraglia, D. Fruchart, J. Marcus and Ph. Monod: cond-mat/0105592.
\bibitem{Nai} Yu. G. Naidyuk, I. K. Yanson, L. V. Tyutrina, N. L. Bobrov, P. N. Chubov, W. N. Kang, H. Kim, E. Choi and S. Lee: cond-mat/0112452.
\bibitem{Kortus} J. Kortus, I. I. Mazin, K. D. Belashchenko, V. P. Antropov and L. L. Boyer: Phys. Rev. Lett. {\bf 86} (2001) 4656.
\bibitem{Shul} H. Shul, B. T. Matthias and L. R. Walker: Phys. Rev. Lett. {\bf 15} (1959) 552.
\bibitem{Kondo} J. Kondo: Prog. Theor. Phys. {\bf 29} (1963) 1.
\bibitem{Soda} T. Soda and Y. Wada: Prog. Theor. Phys. {\bf 36} (1966) 1111.
\bibitem{Legget} A. J. Legget: Prog. Thoer. Phys. {\bf 36} (1966) 901.
\bibitem{Gusman} G. Gusman: J. Phys. Chem. Solids {\bf 28} (1967) 2327.
\bibitem{Chow1} W. S. Chow: Phys. Rev. {\bf 172} (1968) 467; {\it ibid}. {\bf 176} (1968) 525; {\it ibid}. {\bf 179} (1969) 444; Phys. Rev. B {\bf 3} (1971) 1773; {\it ibid}. {\bf 5} (1972) 4404.
\bibitem{Chow2} R. H. Burkel and W. S. Chow: Phys. Rev. B {\bf 3} (1971) 779.
\bibitem{Tang} I. Tang: Phys. Rev. B {\bf 8} (1973) 1276.
\bibitem{Sung} C. C. Sung and L. L. Lacy: Phys. Rev. B {\bf 9} (1974) 2929.
\bibitem{Shiba} H. Shiba: Prog. Theor. Phys. {\bf 40} (1968) 435.
\bibitem{foot1} The self-energy correction also affects the kinetic term. However, this effect can be absorbed into the chemical potential as far as the one-particle properties, such as the density of states, are considered. 
\bibitem{Maki} K. Maki, in {\it Superconductivity} ed. R. D. Parks (Marcel Dekker, N.Y.,1969) Vol.2, p.1035.
\bibitem{note3} Although the order parameters $\Delta_a$ and $\Delta_b$ are more fundamental quantities than ${\tilde \Delta}_a$ and ${\tilde \Delta}_b$, we discuss ${\tilde \Delta}_a$ and ${\tilde \Delta}_b$ because they are directly related to the energy gap in the superconducting density of states.
\end{thebibliography}
\end{document}